# Origin of giant photocontraction in obliquely deposited amorphous $Ge_xSe_{1-x}$ thin- films and the intermediate phase

Mingji Jin, Ping Chen, P. Boolchand

*Department of ECECS, University of Cincinnati, Cincinnati, OH 45221, USA*

T. Rajagopalan, K. L. Chopra

*Indian Institute of Technology, Hauz Khas, New Delhi -1100016, India*

K. Starbova and N. Starbov

*Central Laboratory of Photoprocesses 'Acad J. Malinowski"*

*Bulgarian Academy of Sciences, Acad. G. Bonchev St., Bl 109*

*1113 Sofia, Bulgaria*

## Abstract

Obliquely deposited amorphous $Ge_xSe_{100-x}$ thin-films at several compositions in the 15% < x < 33.3% range , and at several obliqueness angles in the 0 < α < 80° range at each x were evaporated on Si and glass substrates. Here α designates the angle between film normal and direction of vapor transport. Raman scattering, ir reflectance and optical absorption measurements were undertaken to characterize the vibrational density of states and optical band gaps. Edge views of films in SEM confirm the columnar structure of obliquely (α = 80°) deposited films. Films, mounted in a cold stage flushed with $N_2$ gas, were irradiated to UV radiation from a Hg-Xe arc lamp, and photo-contraction (PC) of oblique films examined. Compositional trend of PC exhibit a bell shaped curve with a rather large effect ( 25%) centered in the 20 %< x < 25% range, the Intermediate Phase (IP) with the PC decreasing at x > 25% , the stressed-rigid phase, and at x < 20%, the flexible phase. IR reflectance confirmed absence of photo-oxidation of films under these conditions. The IP represents a range of compositions across which stress-free networks form. Columns observed in SEM reveal a high aspect ratio, with typical lengths in the 1-2 μm range and a lateral width in the 50 nm range. We observe a blue shift (up to 0.38 eV) in the optical bandgap of oblique films (α = 80°) in relation to normally deposited (α = 0°) ones, a result we identify with carrier confinement in nano-filaments ( < 10 nm), that form part of columns observed in SEM. In the IP, the large PC results due to the intrinsically stress-free character of filaments, which undergo facile photomelting resulting in film densification. Ge-rich films ( 25% < x < 33.3%) are intrinsically nanoscale phase separated, and consist of nano-filaments ( ~$Ge_{25}Se_{75}$) that demix from a Ge-rich ( ~$Ge_{40}Se_{60}$) phase that fills the inter-columnar space. Loss of PC in such films is traced to the growth of the Ge-rich phase, which is stressed and photo-inactive. In contrast, Se-rich films are homogeneous, and loss of PC as x decreases below 20% is traced to the accumulation of network stress in the Se-rich nano-filaments. The microscopic origin of the giant PC effect in amorphous semiconducting thin-films can be traced, in general, to three conditions being met (i) growth of a columnar structure leading to porous films, (ii) formation of columns that are *rigid but intrinsically stress-free*, and (iii) an appropriate flux of pair-producing radiation leading to photo-melting of columns. These findings lead us to predict that PC will, in general, be maximized in obliquely deposited films of semiconducting networks glasses residing in their IP, when irradiated with super bandgap radiation.

## I. INTRODUCTION

Vapor deposited amorphous thin-films of the chalcogenides have attracted widespread interest as material systems for information storage [1], light induced effect [2], opto-mechanical effects [3]. When deposited obliquely ( $\alpha > 0$) rather than normally ($\alpha = 0$), these films can become porous because of a columnar structure, and display unusual functionalities [4]. Here $\alpha$ represents the angle between film normal and vapor deposition direction. Starting in the late 1970s, in a series or reports, K. L. Chopra and collaborators noted [9,12,13,14] that obliquely deposited thin-films of amorphous $Ge_xSe_{1-x}$ upon illumination with a Hg-vapor lamp, undergo a giant photo-contraction ( ~20%) in thickness at select compositions. The large photo-contraction (PC) made possible a new lithography for storing relief images [5]. And although columnar structures in obliquely deposited films was observed [6] with many other material systems, the PC effect was found to be selective, and peculiar to certain chalcogenide films only. For example, hydrogenated amorphous Si films when deposited obliquely [7] revealed a columnar structure, but upon visible light illumination showed no PC just degassing. On the other hand an amorphous thin-film of the Ge-As-Se ternary near a mean coordination number ***r*** ~ 2.45 showed one of the largest PC reported to date [8]. Growth of columnar structure is necessary but not a sufficient condition for PC to be manifested.

In 1989, Spence and Elliott [9] examined obliquely deposited amorphous $GeSe_2$ , $GeSe_3$ and $GeS_2$ thin-films, and found Xe-lamp irradiation of these films at ambient environment lead to surface photo-oxidation and band edge shifts. The former was elucidated in IR reflectance measurements and Auger depth profiling, and the later in optical absorption edge shift studies as a function of illumination dose. They found $GeSe_3$, and $GeSe_2$ films irradiated in vacuum or in air to photobleach (increase of bandgap) at low obliqueness angles ( $\alpha = 0°$) but to photodarken (decrease of bandgap) at high obliqueness angles ($\alpha = 80°$). They also found the Tauc optical bandgap of $GeSe_2$ films deposited at normal incidence ($\alpha = 0°$, 2.06 eV) to be higher than those of films deposited at high obliqueness angle ($\alpha = 80°$, 1.84(4) eV).

The view was corroborated by E. Marquez *et al.* [10,] who also found photobleaching of obliquely deposited ($\alpha$= 70) $GeS_2$ films when exposed in air to radiation from a Hg lamp. Their IR transmission experiments also provided evidence of Ge-O bonds serving to confirm surface oxidation of films when illuminated in air. The morphology, hardness and optical properties of obliquely deposited $As_2S_3$, $GeS_2$ and AgBr films were studied by I. Starbova and collaborators [11], who confirmed growth of columns and found a steady decrease in hardness of films with obliqueness angle ($\alpha$).

Starting in the late 1990s, one has discovered the existence of compositional windows[12], [13, 14, 15, 16] in binary and ternary chalcogenide glasses across which glass transitions become almost completely reversing. The non-reversing enthalpy associated with the glass transition ($T_g$) is found to nearly vanish [13, 17]. Raman scattering [13, 14] through measurements of optical elasticity power-laws ( see below) has also shown that glass compositions in these windows form networks that are *rigid but stress-free*. Such networks do not age much, as revealed by the lack of changes [14] in the non-reversing enthalpy associated with $T_g$ as a function of waiting time for periods extending up to years. These privileged compositions are identified as forming part a novel elastic phase called the *Intermediate Phase* (IP) [18]. Experiments have also shown that such networks when illuminated to near band-gap radiation, energy from the photon field is found to strongly couple to valence electrons leading to profound photostructural transformations including giant photo-acoustic softening[19] and photo-melting[20]. Such is not the case for amorphous or glassy network compositions residing outside the IP. The rigid but unstressed nature of networks formed in the IP, as we shall show here, is an important feature for PC effects to be manifested.

In the present work we have measured the PC effect in obliquely deposited $Ge_xSe_{1-x}$ thin-films, and find the PC effect to be globally maximized in the IP[18]. One finds PC to decrease as one goes away from the IP; both above that phase at $x > 26\%$ , and below that phase at $x < 20\%$. The result opens for the first time a way to understand the microscopic origin of PC effects at a basic level. The molecular structure and optical band gap of films at several compositions 'x', are examined as a function of obliqueness angle respectively in Raman scattering and spectrophotometric measurements. These data have permitted us to develop a structural model of these films to understand the molecular origin of the PC effect. Our results reveal that giant PC effect in amorphous semiconducting thin-films can be traced, in general, to three conditions be satisfied , (i) growth of a columnar structure leading to porous films, (ii) columns formed in such films possess intrinsically a rigid but stress-free network structure, and (iii) films when irradiated to an appropriate flux of pair-producing radiation, leads energy from the photon field to be rapidly transferred to valence electron[4] leading to densification of films as columns photo-melt. These three conditions must all be met for giant PC effects to be observed. These findings lead us to predict that PC will, in general, be maximized in obliquely deposited films residing in the IP of chalcogenide glasses.

The paper is organized as follows. In section II we introduce general ideas on elastic phases of network glasses and comment on the IP of binary $Ge_xSe_{1-x}$ glasses of interest here. In section III we present experimental procedure and results. These results are discussed in section IV. A summary of the present findings is presented in section V.

## II. INTERMEDIATE PHASE IN BULK $Ge_xSe_{1-x}$ GLASSES

. There is growing evidence to suggest that the *physical behavior* of network glasses is captured in their elastic behavior [21, 22, 23]. Network glasses are generally of three kinds, elastically *flexible*, elastically *rigid but unstressed* (IP), and elastically *rigid but stressed*. *Flexible* glasses usually form in networks that are weakly connected such as chains of Se, where every atom has 2 near neighbors or the mean coordination number $\boldsymbol{r}$ = 2. On the other hand, *stressed rigid* glasses usually form when networks are highly cross-linked, i.e., when $\boldsymbol{r} \sim 3$. *Rigid but unstressed* glasses usually form when networks possess an optimal connectivity, usually in the $2.3 < \boldsymbol{r} < 2.5$ range. The term optimal here corresponds to the circumstance when the Lagrangian constraints associated with bond-stretching and bonding forces equal the 3 degrees of freedom per atom [21]. These three regimes of elastic behavior span network connectivity in the $2 < r < 3$ range usually, and have been observed in more than a dozen families of covalent glasses such as chalcogenides and oxides and are reviewed elsewhere[24, 25].

What do we know about the IP in binary $Ge_xSe_{1-x}$ glasses? The IP in such glasses was established [20] from Raman and modulated DSC experiments to reside in the 20% < x < 25% range, and the stress free character of this phase was demonstrated in Pressure Raman experiments [26]. Glass compositions in the IP appear to be functionally quite distinct from those outside this compositional window, both below ( x < 20%) and those above (x > 25%) it. For example, we have already alluded to the fact that IP glasses do not age much, a functionality that must be the consequence of the quasi-equilibrium nature of these networks that are in a stress-free state. Furthermore, there is evidence to show that near-band gap radiation strongly couples to such networks as revealed in Raman scattering [20] in Brillouin scattering [19] experiments. In Raman scattering, one observes vibrational modes of corner-sharing (CS) and edge sharing (ES) tetrahedra in these glasses. These vibrational modes are reasonably narrow and also well resolved in the lineshape, and permit one to accurately measure their frequencies to 1 part in a thousand. The square of the mode frequency of CS tetrahedral, $\upsilon_{CS}^2$, serves as a useful measure of network optical elasticity. It increases systematically as network connectivity increases upon cross-linking. Variations in mode frequency of CS- tetrahedra, examined systematically as a function of composition x, display a characteristic optical elasticity power-law in the IP that is distinct from the one observed in stressed rigid glasses [20]. On the other hand, when these Raman experiments are performed at high (factor of $10^3$) laser power-densities, such as in micro-Raman measurements (laser light brought to a fine focus, spot size 1μm) instead of macro-Raman experiments (laser light bought to a loose focus, spot size 50 μm) , one finds the IP to collapse [20]. At high power-densities, a rapid bond switching of Ge-Se, and Se-Se covalent bonds occurs [27], and the " cooperatively self-organized network structure"  is optically driven to a "random network structure".  In the optically pumped state, glass network structure melts and leads to a collapse of the IP[20]. These observations underscore that pair-producing radiation can facilely couple to disordered networks when they are in a rigid but stress-free state. In summary, the observation of a characteristic elastic power-law, the lack of aging and thermal reversibility of $T_g$s , the *effective* transfer of energy from pair-producing  photon field to covalent bonds in such networks, are viewed as consequences of the *rigid and stress-free character* of networks formed in the IP. These functionalities observed in IPs of inorganic glassy networks share commonalities with special biological networks called Proteins [28], and are broadly identified with 'self-organization' of these disordered networks.[16]  These ideas will have a bearing on the PC effects that we will report here.

## III. EXPERIMENTAL

In this section we provide details of thin-film growth and characterization of these films by Raman scattering, Scanning Electron Microscopy, IR reflectance and Optical absorption studies.  Details on measurements of the Photo-contraction effect using profilometry and an illumination facility in an inert ambient will also be described.

### A. $Ge_xSe_{1-x}$ thin film synthesis

Bulk $Ge_xSe_{1-x}$ glass samples at  x = 0.15, 0.20, 0.22, 0.23, 0.27, 0.30 and 0.33 were synthesized using 99.9999% elemental Ge and Se as the starting materials, sealed in evacuated (< 5 × $10^{-7}$ Torr) silica ampoules in the desired molar ratio. Melts were homogenized at 1000 °C for at least 48 hours and then equilibrated at about 50 °C above the liquidus for an additional 24 hours before quenching in water. Using these bulk samples as evaporation charges, thin films were deposited on glass and Si substrates by using two set ups. In one set up, used at the IIT in New Delhi, a spherical framework was constructed[29] to support substrates at 5 obliqueness angles $\alpha = 0°, 20°, 45°, 60°$ and $80°$. Films at these angles were deposited simultaneously from a single evaporation charge. Typical operating conditions included a base pressure of $P_{base} = 5 \times 10^{-6}$ Torr and a deposition rate of 2nm/ sec. Films at 5 compositions ( $x = 0.15, 0.20, 0.22, 0.23$ and $0.33$) were prepared in this fashion, and the typical thickness of films was in the $1.6\ \mu m < t < 3.5\ \mu m$ range. In the second set up, used in laboratory of the Academy of Science in Sofia, Bulgaria, films were vapor deposited individually at the desired obliqueness angle $\alpha$ using bulk glass compositions as described above. Bulk glasses used as evaporation charges for both set of depositions were synthesized at University of Cincinnati. In the second set up films were synthesized at $x = 23\%, 27\%, 30\%$ and $33\%$. Typical film thickness ranged from 0.6 μm to 3.2 μm.

Growth of amorphous thin-films from multicomponent chalcogenide evaporation charges is as much an art as a science [30], and it requires tuning the parameter space [31] to achieve homogeneous films of a stoichiometry close to that of the evaporation charge. This becomes a challenge when different elements of widely varying melting points are used. A comparison of the Raman spectra of the bulk glass used as the evaporation charge with the normally deposited ( $\alpha = 0°$) films, (section IIIC below) provides a good check on film quality. Both the groups in New Delhi [29, 8, 4, 32] and in Sofia [11, 33-35] have had prior experience in depositing and characterizing obliquely deposited thin-films.

## B. Photo contraction (PC) measurements on obliquely deposited amorphous $Ge_xSe_{1-x}$ thin- films

Fig. 1 gives a view of the experimental setup used for photo-illumination of thin-films. Thin films were mounted in a hot stage (Model #50-600, Creative Devices Corp.) that was continuously purged with dry $N_2$ gas, and the cell assembly cooled by continuous flow of water. The cell has a ¾ inch glass window permitting radiation from the Xe-Hg arc lamp to irradiate thin-film samples mounted in the cold stage. Use of the cold stage suppressed surface oxidation of films as was confirmed by FTIR measurements (see below). Water cooling of the stage insured that film temperature did not increase during irradiation. Cell temperature was continuously monitored and remained at room temperature during illumination. A 1000W Hg-Xe arc lamp, (Oriel model #6295) with a condenser, a neutral density filter, and a water filter (absorb IR radiation) provided a light beam of 1cm diameter. Sharp edge masks were used to define exposed areas on films for photocontraction measurements. A typical irradiation used light power density of 75 $mW/cm^2$ and lasted for 2 hours. Under these conditions the PC effect is known to saturate as noted earlier [29].

The PC effect on $Ge_xSe_{1-x}$ films deposited at high obliqueness angle $\alpha = 80°$ films was examined as a function of glass chemical composition x. Thickness of virgin and light-illuminated films were measured using a KLA Tencor P-10 surface profiler to an accuracy of a few Å scale. Film thickness reduction ($\Delta t$) could be measured directly in such films of thickness t of 1-3 μm. In our experiments we found convenient to plot directly the observed change in thickness of films as a function of composition.

These results are summarized in Figure 2.

Our results show the PC effect is maximized near x = 22% and 23% and steadily decreases at x > 25% and at x < 20%. Variations in the PC effect are compared to those in the non-reversing enthalpy $\Delta H_{nr}$ near $T_g$ of the corresponding bulk glasses taken from earlier work [26, 36]. One finds a close correlation between the PC effect and the global minimum in the $\Delta H_{nr}$ term; the PC effect is maximized in the IP, and it steadily decreases at x > 25% in the *stressed rigid* phase, and also at x < 20% in the *flexible* phase. Earlier reports of the PC effect while not as complete show similar trends [8]. We shall return to discuss these results in the next section. [36]

### C. $Ge_xSe_{1-x}$ thin-film characterization

**Scanning Electron Microscopy.** Cross-sectional images of select thin-films were examined by a Hitachi S-4000 Scanning Electron Microscope (SEM). Films were mounted in a sample chamber evacuated to $10^{-5}$ Torr , and an electron beam voltage of 20kV and current 184 μA was used to image films. Fig. 3 shows an example of a cross-sectional view of a normally ($\alpha = 0^{o}$) in the (top panel) and obliquely ($\alpha = 80^{o}$) deposited (bottom panel) $GeSe_2$ film on a Si substrate. For the obliquely deposited ($\alpha = 80^{o}$) film , there is evidence of texture in the form of columns about 50 nm in lateral size and running at approximately an angle $\beta = 55^{o}$ with respect to the film normal. Leamy and Dirks [6] had earlier suggested the existence of an empirical rule,

$$Tan\alpha = 2 * Tan\beta \quad (1)$$

where α and β represent respectively the obliqueness angle of deposition and the angle made by the columns with respect to film normal. For the case $\alpha = 80^{o}$, equation (1) yields a value of β = 70.5° reasonably close to the observed value. The columns possess a high aspect ratio typically 1 to 2 μm long and about 50 nm in lateral width. The width is much too large to be ascribed to atomic $GeSe_4$ chains. More likely the columns represent ribbons or bundle of several interlaced atomic $GeSe_4$ chains (see below). The suggestion was made earlier in a theory paper [37] to account for the PC effect.

**Raman Scattering**. Raman scattering measurements used a T64000 triple monochrometer system from Horiba Jobin-Yvon , equipped with a CCD detector and a BX 41 confocal microscope attachment. The scattering was excited using the 647.1 nm line from a $Kr^+$ ion laser. Films were examined in their pristine state. Measurements were performed at room temperature and used a laser power density of typically 0.4 mW/2 $\mu m^2$.Raman lineshapes were deconvoluted using a "Grams/AI" software. We will first begin by presenting Raman data on the as deposited thin-films. Next we will proceed to provide Raman data on select films in the photo-contracted state.

Figures 4, 5 and 6 give Raman lineshapes of films respectively at compositions x = 33%, 30% and 23%, examined as a function of the obliqueness angle α. In each figure we also provide Raman scattering of the corresponding bulk glass, the starting material used to evaporate films. We find that spectra of the normally deposited films, i.e., α = 0 are actually quite similar to those of bulk glasses. These data are a litmus test of film quality. In earlier reports, see for example [31], Raman spectra $Ge_xSe_{1-x}$ films synthesized by evaporation and PECVD were compared to those of bulk glasses. These authors noted that the observed lineshapes of thin-films varied noticeably from those of bulk glasses.

Obliquely deposited $GeSe_2$ films display rather spectacular changes in Raman lineshapes as a function of obliqueness angle α (Figure 4). In these data there are four modes of interest whose microscopic origin has been widely accepted [12, 38, 26] ; a mode near 178 $cm^{-1}$ is ascribed to ethanelike $Ge_2Se_6$ units, a mode near 200 $cm^{-1}$ is that of corner-sharing(CS) $GeSe_{(1/2)4}$ tetrahedra, a mode near 217 $cm^{-1}$ belongs to edge-sharing (ES) $GeSe_2$ tetrahedra, and a mode near 255 $cm^{-1}$ is a stretching vibration of $Se_n$ chain segments. We have deconvoluted the observed lineshape in terms of several Gaussian profiles and have extracted the mode frequency, mode width and integrated intensity. In figure 7a, b and c , we plot respectively the variation in the CS mode frequency ( $\upsilon_{CS}$), ES mode frequency ($\upsilon_{ES}$) and the ratio of the ES to the CS mode scattering strength ( $I(A_1^c)/ I(A_1)$) as a function of α. At x = 33.3% (Figure 4), one observes the $Se_n$ chain mode to grow in intensity, the scattering strength ratio of ES to CS tetrahedra to decrease, and respective mode frequencies $\upsilon_{CS}$ and $\upsilon_{ES}$ to decrease as α increases to 80°. The Raman spectrum of the obliquely deposited film at α = 80° bears a close analogy to that of a bulk glass of composition x ~ 25%. This can be seen by projecting the measured parameters of the films on the plot of Figure 8, which summarizes how these parameters change in bulk $Ge_xSe_{1-x}$ glasses [12 26 38 20] as their Ge content is varied. At x = 30% (Figure 5), the observed lineshapes reveal the $Se_n$ chain mode to increase in intensity and the ratio ( $I(A_1^c)/ I(A_1)$) to decrease in magnitude as α increases to 80°. The behavior of films at x = 30% parallels that found at x = 33.3% (Figure 7), and in particular the Raman lineshapes of films at α = 80°, at both these compositions look surprisingly similar to those of a $Ge_{25}Se_{75}$ bulk glass. The data of Figure 4 and 8 serve as a key in decoding the structural changes taking place in oblique films as a function of α . The red-shift in $\upsilon_{CS}(\alpha)$, and $\upsilon_{CS}(\alpha)$ and the concomitant reduction in the scattering strength ratio of the ES- to CS-mode, $I(A_1^c)/ I(A_1)$underscores that a Se-rich phase of nearly $GeSe_4$ stoichiometry is formed in the obliquely deposited $GeSe_2$ films. These data constitute signature of nanoscale phase separation of films at x > 25%, as we will show in the next section.

The Raman lineshapes of films at x = 23% as a function of α appear in Figure 6, and demonstrate unequivocally that there are little or no changes as a function of increasing α. A perusal of the lineshape and a detailed analysis of these data confirm the finding. In figure 7, variations in the parameters $\upsilon_{CS}(\alpha)$, $\upsilon_{CS}(\alpha)$ and $I(A_1^c)/ I (A_1)$ as a function of α for films at x = 23% show little variation with α. These data confirm quantitatively that $Ge_{23}Se_{77}$ films are, indeed, quite special; their molecular structure does not change appreciably with α even though morphologically these films steadily evolve from a 3D-like structure at α = 0° to a quasi 1D-like structure as columns are manifested at the highest obliqueness angle, α = 80°. The result stands in sharp contrast to that found above for the films at x > 25%.

Films at x = 20% and 15% show Raman lineshapes that do not change much with obliqueness angle. Analysis of the observed lineshapes reveals that variations in parameters such as $\upsilon_{CS}(\alpha)$, $\upsilon_{CS}(\alpha)$ and $I(A_1^c)/ I (A_1)$, and particularly $\upsilon_{Se}(\alpha)$ and $I(Se_n)/I(A_1)$ suggests that these films are becoming steadily Se-richer as a function of α Figure 7. These data suggest that the excess Se is being incorporated in the columns, a point we return to discuss later.

The confocal feature of the Olympus BX 41 microscope has permitted us to record Raman spectra of obliquely deposited films (α = 80°) at x = 33.3% as a function of depth from film surface as illustrated in Figure 9. We examined a photocontracted film with a virgin as deposited film. We note that as we go deeper in the film to a depth d = 5a, the virgin and photoilluminated film lineshapes become indistinguishable. Here "a" represents a full rotation of the knob controlling the z-displacement. We estimate a to represent a depth of 0.2 μm. These data suggest that the exciting radiation from the Hg-Xe arc lamp is apparently absorbed on film surface. As we show in the next section such films possess an optical gap of 2.70 eV, and the relevant exciting radiation in this case is the 435 nm or 2.84 eV light from the Hg-Xe lamp that partakes in PC. In this particular case, the PC effect largely occurs on the film surface. Such is, however, not the case for other films that possess a smaller bandgap and the exciting radiation is most likely the 546 nm or 2.27 eV light from the Hg-Xe lamp source.

**Optical bandgaps.** Optical band gaps of the amorphous $Ge_xSe_{1-x}$ thin-films provide valuable complimentary information to Raman scattering on the molecular structure of films. Bandgaps of films in their *as deposited state* on glass substrates were determined using a Shimadzu 2501PC UV-VIS spectrophotometer. The optical absorbance was measured in the 400-700 nm range using a 1nm slit width, and the data were normalized to bare glass as a reference. Figure 10 gives a summary of results on films deposited normally ($\alpha = 0°$). From the measured film thickness and absorbance, we deduced the absorption coefficient $\mu(E)$. Variations in the absorption coefficient $\mu$ (E) as a function of photon energy appear in Figure.10 (b). The optical band gap, of films corresponding to an absorption coefficient $\mu = 5 \times 10^3$ cm$^{-1}$ , henceforth denoted as $E_{53}$, was established, and Figure.11(a) gives a summary of the results. We find the optical band gaps of normally deposited films systematically increase with x. Earlier Street *et al.* [39] reported bandgaps ($E_{13}$) of evaporated $Ge_xSe_{1-x}$ thin-films corresponding to an absorption coefficient $\mu = 1 \times 10^3$ cm$^{-1}$. Their data are compared to the present data in Figure 11. We find that these two data sets generally follow the same trend, namely an increase in the bandgap with Ge content of films. Here we must mention that the $E_{13}$ gaps reported are those on thermally annealed films. In our experiments we did not investigate changes in bandgap upon thermally annealing films.

We have also deduced the Tauc band gap ($E_T$) of our films by plotting $(\mu h\omega)^{1/2}$ against photon energy E (= hω) and from the observed linear variation extrapolated the energy E = $E_T$ where μhω vanishes. Fig.10 (c) gives a plot of $(\mu h\omega)^{1/2}$ against E for films investigated by us. The Tauc band gap ($E_T(x)$) , summarized in Figure 11(b), show a rather systematic upshift as the Ge content of films increases. On this plot we have also projected the Tauc bandgap of $GeSe_2$ films reported by Spence and Elliott[9] in the as deposited (2.06 eV) and after optical irradiation (2.13 eV) state. These Tauc bandgap values for $GeSe_2$ are lower than the values observed in our films.

How do the bandgaps change in obliquely deposited films? In Figures 12, 13 and 14, we provide respectively data for films at x = 22%, 23%, and 30%. In each figure there are 4 panels; (a) plots the measured absorbance as a function of photon energy (E), (b) variation of the absorption coefficient μ(E), (c) plots of $(\mu h\omega)^{1/2}$ against E, and (d) variations of the $E_{53}(\alpha)$ bandgap and the $E_T(\alpha)$ Tauc bandgap as a function of the angle α. A perusal of these data show that $E_{53}(\alpha)$ bandgaps generally increase with α but not monotonically. For compositions at x = 22%, 23%, 27%, the bandgaps at first decrease as α increases from 0° to 40° or 60°, and thereafter one observes the gap to increase as α increases to 80°. The threshold behavior nearly vanishes as x increases to 30% (Figure 14) and bandgaps increase almost monotonically with α.

In figure 15 and Table 1, we compare the $E_{53}(\alpha)$ bandgaps of normally deposited ( α = 0) with obliquely deposited films ( α = 80). We find that the bandgaps of obliquely deposited films are significantly larger than of normally deposited films. At x = 33.33%, the upshift in the energy gap is nearly 0.38 eV, and it steadily decreases to nearly vanish as x decreases to 15%. The result for films at x = 33.33% is all the more interesting given that our Raman results reveal obliquely deposited $GeSe_2$ films display a lineshape characteristic of a film of $Ge_{25}Se_{75}$ composition. There are clearly other factors that come into play to understand these data, issues we discuss in the next section.

**IR reflectance.** IR response of thin-film samples in the virgin and photocontracted state were measured using a Thermo Nicolet FTIR spectrometer model 870 using a variable angle reflection attachment called Seagull marketed by Harrick Scientific Products, Inc. DTGS with polyethylene or KBr window as detector, with KBr and solid substrate used as beam splitters, permitted the IR response from the films to be recorded in the Far and Mid IR regions. The instrument is supported by OMNIC and Grams/AI software to acquire and analyze data. In a typical measurement, 200 scans were programmed to yield 4 $cm^{-1}$ resolution and the response over a wide frequency range from 200 to 1200 $cm^{-1}$ was recorded. We also recorded ir response from a $GeO_2$ glass, and as expected observe vibrational features of tetrahedrally coordinated Ge to oxygen neighbours near 900 $cm^{-1}$, as illustrated in Figure 16(a) and (b). We confirm that films become partially oxidized if the cold stage is not thoroughly purged with $N_2$ gas during irradiation from the Hg-Xe lamp. We illustrate the finding in Figure 16(b), which compares the ir response of a film at x = 23%, both before and after irradiation. On the other hand, in the presence of a $N_2$ gas flow, we found no evidence of oxidation for a film at x = 22% as illustrated in Figure 16(a). Thus, we believe that the observed trends in PC, band-gap variation, and Raman scattering reported here on amorphous Ge-Se thin-films represent the intrinsic behavior of these films, and are not related to any surface oxidation effects [9].

## IV. DISCUSSION

### A. Nanoscale phase separation of obliquely deposited $Ge_xSe_{1-x}$ thin-films

The Raman scattering of obliquely ( α = 80°) deposited $GeSe_2$ films reveal lineshapes that show a close similarity to a glass of $GeSe_3$ stoichiometry ( Figure 4,7). On the other hand the bandgap of such films (2.68 eV) is considerably larger than the gap of $GeSe_3$ glass (2.38 eV). How are we to reconcile these results? We would like to suggest that these films are intrinsically segregated on a nanoscale into two distinct phases, one of these (A) consist of columns of nearly $GeSe_3$, stoichiometry and the other phase (B) is of nearly $Ge_2Se_3$ stoichiometry. Nanophase B localizes in the inter-columnar regions, and the underlying nanoscale phase separation can then be described by the following equation:

$$Ge_{1/3}Se_{2/3} = \left(\frac{4}{9}\right)Ge_{1/4}Se_{3/4}(A - phase) + \left(\frac{5}{9}\right)Ge_{2/5}Se_{3/5}(B - phase) \qquad (2)$$

The first term on the right hand side of equation 2 represents the Se-rich A nano-phase while the second term the Ge-rich phase B-nanophase. Equation (2) suggests that the concentration of these two nanophases must be comparable ( 4:5). In practice Raman scattering is dominated by the A nanophase because its bandgap is close to that of the 647 nm (1.96 eV) exciting radiation. Indeed, one can account for the observed lineshape in terms of contributions of these two nanophases as shown in Figure 17. The lineshapes of the two nanophases are taken from those of corresponding bulk glasses. Films at x = 23% reveal Raman lineshapes that are *independent* of the obliqueness angle. Oblique films at this privileged composition consist only of the A nanophase present as columns. Raman lineshape parameters, $\upsilon_{CS}(\alpha)$, $\upsilon_{CS}(\alpha)$ and $A_{ES}/A_{CS}(\alpha)$, of films at x = 23% place the stoichiometry of the nanophase A close to $GeSe_3$. These Raman lineshape parameters in bulk $Ge_xSe_{1-x}$ glasses have been established in the 15% < x < 40% range at every 2 at.% interval (figure 8). The present data on films shown in Figure 7 can thus be analyzed to provide details of film molecular structure. It appears that nanophase A can exist in either a 3D morphology as in a bulk glass (Figure 8) or in quasi-1D columns as in thin-films deposited at high obliqueness angle α= 80° films (Figure 6).

Our Raman data suggest a totally different picture for thin-films deposited at low x, i.e., x =23% and 20%. In Figure 7, we show a plot of the three parameters, $\upsilon_{CS}(\alpha)$, $\upsilon_{CS}(\alpha)$ and $I_{ES}/I_{CS}(\alpha)$, for films at x = 20% and 23% and these parameters remain largely independent of obliqueness angle. These lineshape parameters serve to define the network backbone, and suggest that the chemical stoichiometry of films remains independent of α. Even when columns appear at high obliqueness angle (α = 80 degrees), their stoichiometry is close to that of bulk glass used as the starting material.

For the Se-rich film at x = 15%,a perusal of the data of Fig.7 shows that the three parameters, $\upsilon_{CS}(\alpha)$, $\upsilon_{CS}(\alpha)$ and $I_{ES}/I_{CS}(\alpha)$, are all larger than those for films at x = 20% and 23%. In particular, the mode frequencies of CS and ES tetrahedral units are present in a local environment that is quite similar to that of a bulk glass near x = 25%. These data suggest that films at x = 15% are also intrinsically heterogeneous - composed of Ge-rich regions and Se-rich regions. The three parameters above serve to define how Ge atoms bond in films. On the other hand, information on the Se-rich phase comes more reliably from an analysis of the $Se_n$ chain mode frequency, and particularly the ratio R of the scattering strength of the broad $Se_n$ mode relative to the CS mode. Our Raman results reveal the ratio R to increase from 1.1 at $\alpha = 0°$ to 1.7 at $\alpha = 80°$. These data suggest that at low obliqueness angles films are intrinsically segregated while at high obliqueness angles, as columns appear, they are representative of a network with a stoichiometry of about x = 15%.

Molecular structure results on films suggested from the Raman data can be briefly summarized. Films at x > 25% are intrinsically segregated on a nanoscale into two phases A and B, with the A-phase representing the column material and the B-phase the inter-columnar material. Films at x = 20% and 23% are intrinsically homogeneous, i.e., there is no evidence of a variation in stoichiometry with α. On the other hand, film at x = 15% appears to be again segregated into Ge-rich and Se-rich regions, and such segregation almost vanishes as α increases to 80°.

## B. Bandgap variation with Oblique deposition and Carrier Localization

Earlier work by Street *et al.* [39] and P. Nagels *et al.* [31], had established that the optical gap of normally deposited $Ge_xSe_{1-x}$ thin-films increases with Ge concentration (Figure 11). Our results confirm the finding, and reveal optical gaps of films to be larger than those reported earlier at the same film stoichiometry x. In the present work, films in most cases, were relaxed at room temperature for over a year before examination. Raman data on normally deposited films are found to be quite similar to those of the bulk glasses used as starting materials to evaporate them. These data suggest that our films are homogeneous and relaxed, and the larger bandgap most likely a consequence of these circumstances.

The optical bandgap of our thin-films examined as a function of obliqueness angle α display interesting trends. For film compositions close to the threshold value , x ~22%, bandgaps are found ( Figure 12d) to first decrease with α , to show a broad and shallow minimum in the 45 °< α < 60° range, and to increase thereafter to show a maximum at high obliqueness angle, α = 80°. The behavior is also observed at x = 23% (Figure 13d) . But as x increases to 30% , the broad and shallow minimum in the bandgap becomes less conspicuous, and a general increase of the bandgap is observed with α (Figure 14d) . In Figure 15, we compare the optical gap of our normally deposited films (α =0) with those of obliquely deposited (α = 80) ones as a function of film stoichiometry x. The results show that both gaps increase with x, however the increase in the gap of obliquely deposited films, particularly at x > 23% is significantly larger than of normally deposited films. In particular, $E_{53}$ bandgap of α =80 films (2.68 eV) (Figure 15) is found to be much larger than the gap of normally deposited ( α =0) films ( 2.30 eV).

The dielectric constant (ε) of $GeSe_2$ glass was established from IR reflectance measurements [40] and found to have a value of ε = 12.31. If one assumes the electron effective mass, $m^*$, in these films to be 0.1m, a value typical of many compound semiconductors [41], we estimate the exciton radius R in amorphous $GeSe_2$ films to be,

$$R = a_0\left(\frac{m^*}{m}\right)\varepsilon = 6.3nm \qquad (3)$$

In equation (3), $a_o$ represents the first Bohr radius of 0.051 nm and m the rest mass of an electron. The blue shift of the $E_{53}$ optical gap, we would suggest represents localization of carriers in the columns. This can only happen if the columns are of nanoscale dimensions, i.e., of the order of R ~ 10 nm or so. We are, thus, led to the notion that carriers must be localized in nanometric sized atomic filaments of $GeSe_3$. These atomic filaments of $GeSe_3$ are probably made up of chains of corner- sharing $(GeSe_{1/2})_4$ tetrahedra that are bridged by edge sharing $Ge(Se_{1/2})_4$ ones, the structural motif found in the crystal structures of the high temperature form of $GeSe_2$ and the metastable crystalline phase of $Ge_4Se_9$ [42]. The columns observed in SEM investigations (Figure 3) of about 50 nm diameter in lateral dimension, must then be viewed as bundles[37] of nanometric sized atomic filaments, and the 0.38 eV blue shift in bandgap of such films is the result of confinement of electron –hole pairs in such nano-filaments. The data of Figure 15 reveals that such a carrier confinement effect is present in films at $x > 25\%$, and that the effect increases monotonically as x increases to 33.3%. Carrier localization effect in nanophase A is accentuated by the presence of B nanophase. The latter grows as x increases and it serves to laterally confine of the nano-filaments (A). One can then think of an obliquely deposited $\alpha = 80°$ film at $x = 33.3\%$ as composed of multilayers of alternating $GeSe_3$ nano-filaments (A) and $Ge_{40}Se_{60}$ layers (B) growing at an angle $\beta \sim 70°$ (see equation 1). To summarize, the large increase in bandgap of obliquely deposited films in relation to normally deposited ones, we believe, results due to carrier localization effects in nano-filaments that form bundles and are observed as columns in SEM images. We believe that the chemistry of the columns remains close to $GeSe_3$ in all films at high x (> 25%). A glass network of $GeSe_3$ stoichiometry will reside in the IP of the Ge-Se binary and would form stress-free structures.

## C. Optimization of the GPC effect in the Intermediate Phase

It is widely known that porous thin-films of many materials including Si can be grown by oblique deposition. But only a select few of these obliquely deposited films undergo a thickness contraction when illuminated by pair-producing radiation [37]. In the case of Si, for example, little or no photocontraction is observed [7]. Thus, there are factors other than columnar growth that must come into play to account for the selective nature of the PC effect. The optimization of the GPC effect in the IP of the Ge-Se binary (Figure 2) provides an important clue in understanding the molecular origin of the effect.

In the binary $Ge_xSe_{1-x}$ system, it has recently been shown [26] from Raman scattering experiments performed as a function of hydrostatic pressure, that glass compositions in the IP form networks that are stress-free. In these experiments on bulk glasses, one has found the existence of a threshold pressure [38], $P_c$ , that must be exceeded by the applied pressure P for the Raman active vibrational mode of corner-sharing $GeSe_4$ tetrahedra (near 200 $cm^{-1}$) to blue-shift. The threshold pressure, $P_c(x)$, serves as a measure of network stress. It must be exceeded by the applied pressure for the network to be compressed and sense the applied stress. Interestingly, these experiments show that the compositional variation of $P_c(x)$ closely tracks that of the non-reversing enthalpy term $\Delta H_{nr}(x)$ of the glass transition [26]. Glass compositions in the IP ( $20\% < x < 25\%$) display a vanishing $P_c$ and $\Delta H_{nr}$, and both these terms change remarkably as one goes away from the IP, both above ( $x > 25\%$) and below ($x < 20\%$) the IP. The vanishing of $P_c$ in the IP is a remarkable finding. The vanishing of $P_c$ is a result characteristic of a crystalline solid, i.e., a network that is at equilibrium. These data underscore that glass compositions in the IP form networks that are stress-free and in a state of quasi-equilibrium.

The second set of experiments of relevance here bear on the giant photo-acoustic softening observed in binary $Ge_xSe_{1-x}$ glasses in Brillouin scattering measurements [19] when their composition x is close to 20%, the mean-field rigidity transition. The exciting radiation in these Brillouin experiments ( 647 nm radiation from a Kr-ion laser) apparently strongly couples to the network when it is stress free. A parallel behavior is encountered in Raman scattering when the power-density of the exciting radiation ( 647 nm ) is substantially increased [12], and one finds the IP structure to collapse to its centroid near x = 23%. The photo-acoustic softening in Brillouin experiments and the photomelting of the IP in Raman scattering experiments, both performed as a function of laser power-density have at their base the same central idea. Near bandgap light leads to facile photomelting [27] of stress-free disordered networks but not of stressed ones (see below).

The optimization of the GPC effect in films residing in the IP can now be commented upon. We can gauge stress in evaporated films by looking at the widths of Raman vibrational modes. A perusal of these data reveals that at x = 23%, full width at half maximum of the ES mode in obliquely deposited thin-films (α= 80°) is $\Gamma_{ES}^{tf}$ = 10.75 (50)cm$^{-1}$ , nearly the same as in the bulk glasses, $\Gamma_{ES}^{bg}$ = 10.75(50) cm$^{-1}$. At x = 30% , we find $\Gamma_{ES}^{tf}$ = 11.50(50) cm$^{-1}$ to be greater than $\Gamma_{ES}^{bg}$ = 9.70(50) cm$^{-1}$. At x = 33.3%, we find $\Gamma_{ES}^{tf}$ = 16.00(50) cm$^{-1}$ again to be greater than $\Gamma_{ES}^{bg}$ = 9.70(50) cm$^{-1}$ . At x = 15%, we find $\Gamma_{ES}^{tf}$ = 18.00(50) cm$^{-1}$ to be substantially greater than $\Gamma_{ES}^{bg}$ = 10.30(50) $cm^{-1}$. These data reveal that the deposited films at compositions both above and below the IP are inhomogeneously stressed while those in the IP are not.

Near the composition x = 23%, we have a 3D network in the normally deposited ($\alpha$= 0°) film but a quasi 1D columnar structure in the obliquely deposited ( $\alpha$ = 80°) films. In the latter case, films are porous, and the porosity derives from the free volume present between columns. Our Raman scattering results show that the molecular structure of these films is independent of $\alpha$. In both instances ( $\alpha$ = 0° and 80°films) we have networks that are stress-free. When oblique films are exposed to radiation from Hg-Xe lamp, photomelting of the columns occurs, i.e., the columns collapse, and lead to a change in film thickness as films densify. When oblique films are exposed to radiation from Hg-Xe lamp, photomelting of columns occurs, i.e., the columns collapse, and lead to a change in film thickness as films densify.

The underlying photo structural transformations can be described as follows. The minuscule non-reversing enthalpy at $T_g$ ( Figure 2) for networks in the IP suggests that such networks possess high configurational entropies that are close to liquid entropies. Just the reverse is the case for networks residing outside the IP, which possess an order of magnitude larger non-reversing enthalpy of melting at $T_g$ , and must sit at the bottom on an entropy landscape. The role of pair-producing radiation is to create electron- hole pairs, leading to excitons, which eventually recombine delivering band-gap worth of energy (2 eV) to covalent bonds locally, permitting them to switch by a process that has been described by H. Fritzsche Ref. 27. With increasing flux of UV radiation, switching of bonds becomes pervasive and leads to facile light induced melting of networks in the IP since in their pristine state (absence of light) they almost have liquid-like entropies. On the other hand, networks residing outside the IP, which possess a much larger non-reversing heat flow, have a larger entropic barrier to overcome before the photomelted state can be realized. These ideas provide a physical basis to understand how pair producing radiation interacts so differently with stress-free networks than with stressed ones.

At higher x, such as at x = 30% and 33.3 %, oblique films nanoscale phase separate into columns of $GeSe_3$ stoichiometry ( phase A) with the $Ge_{40}Se_{60}$ phase (B) filling pores between columns as schematically illustrated in Figure 18. Here the stress-free A phase is photo-active phase, while the stressed B phase is photo-inactive. The PC effect in oblique films ($\alpha = 80°$) steadily decreases as x increases above 27% because of the presence of the photo-inactive B phase filling the pores. Some reconstruction of the A with B nano phase occurs upon photo-illumination as photomelting proceeds precluding a complete collapse of columns that would have occurred if the B-nanophase were absent. The evidence of A and B nanophases reconstructing upon PC comes from our Raman scattering data examined as a function of depth (figure 9). There are two sets of spectra shown as a function of depth 'd" using a confocal microscope. Here d refers to a depth of about 0.2μm, and corresponds to a complete rotation of the knob affecting a z-displacement of the focal point of the objective. Note that at a depth of 2d the $Se_n$ chain mode near 260 $cm^{-1}$ decreases in strength as the strength of the ES mode near 217 $cm^{-1}$ increases. We visualize these changes as representing the $Ge_{25}Se_{75}$ A phase reconstructing with the $Ge_{40}Se_{60}$ nanophase B to form a $GeSe_2$-like network. The underlying process is the same as given by equation 2 except now the reaction proceeds from right to left, partially undoing the demixing that occurred during oblique deposition of films. These considerations provide a sound basis to understand the reduction of the PC at x > 25% ( Figure 2).

Why does PC effect steadily decrease in oblique films at x < 20%? We have already noted that Raman scattering of films at x < 23%, display vibrational features (Figure 6, 7) that are largely independent of the obliqueness angle. In other words, the chemical stoichiometry of columns formed in oblique films at x < 23%, must be the same as that of the 3D networks formed in normally deposited films. Incorporation of excess Se in the optimally connected $GeSe_3$ columns, renders them progressively stressed. The stressed character of networks in the flexible phase was independently established from the Pressure Raman experiments alluded to earlier[43]. Radiation from the Xe-Hg -lamp now no longer couples well to glass networks in the flexible phase as it did when columns were stress-free at x = 23%. Photo melting of the columns is now suppressed with the result that the PC effect almost vanishes at x = 15% (Figure 2). These considerations on the contrasting role of light interaction with stressed and stress-free networks, provides a basis to understanding why the PC effect is optimized in the IP.

### D. Optical absorption changes upon illumination of obliquely deposited $Ge_xSe_{1-x}$ thin-films

In our experiments we did not investigate optical absorption shifts in obliquely deposited films as a function of light irradiation. Such a study was however reported earlier by Spence and Elliott [9], and Rajgopalan et al [44]. For films irradiated in vacuum, these workers found normally deposited films ( $\alpha = 0°$) to photobleach, while obliquely deposited ones ($\alpha = 80°$) to photodarken. Given our findings of a change in the optical band gap with obliqueness angle (figures 12-14), a rather straightforward interpretation can be advanced to understand the switch from photobleaching to photodarkening as a function of obliqueness angle α. Freshly deposited films are usually stressed, and these can be relaxed either by a thermal anneal or by illumination to near band gap radiation. For Ge-Se films, the optical band edge usually shifts up as stress is released leading to a photobleaching. Films deposited at high obliqueness angles usually have a larger bandgap as found in the present work. Upon illumination to visible light one can expect columns to collapse, and the underlying enhancement of the optical gap due to carrier confinement effects to be lost. The natural consequence will be a lowering of the bandgap, i.e., photodarkening. These stress related effects are not expected to be strongly influenced by surface oxidation of films since these are still largely bulk rather than surface effects[9].

The present findings on the origin of PC in obliquely deposited films of the Ge-Se binary can be formalized into a general prediction. One expects the PC effect to be optimized for obliquely deposited thin-films when their chemical stoichiometry resides in the IP of corresponding bulk glasses. IPs in several families of binary and ternary chalcogenide glass systems has now been identified [25]. For example, in the Ge-As-Se ternary system, a fairly large domain of compositions resides in the IP[45.] In some early work reported on this ternary glass system, one of the largest PC effect (26%) was observed [32] for the $Ge_{0.20}Se_{0.75}As_{0.05}$ composition. This particular composition corresponds to a mean coordination number $\boldsymbol{r}$ = 2.45, if one takes Ge, As and Se to be 4,3 and 2-fold coordinated. The composition resides in the IP [45] of the Ge-As-Se ternary. We predict giant PC effect will also be observed in the ternary Ge-P-Se, binary As-Se, binary P-Se over a wide range of compositions encompassing the IP.

## Summary

Obliquely deposited $Ge_xSe_{1-x}$ thin-films at several compositions in the 15% < x < 33.3% range , and at obliqueness angles in the 0 < α < 80° range at each x were evaporated on Si and glass substrates. Raman scattering, IR reflectance and optical absorption measurements were undertaken to characterize the vibrational density of states and optical band gap of films. Edge views of the films in SEM confirm the columnar structure of the obliquely deposited films [29]. Films mounted in a cold stage flushed with $N_2$ gas, were irradiated to Hg-Xe arc lamp, and confirmed oblique films (α = 80°) to undergo photo-contraction (PC). IR reflectance measurements confirm the absence of photo-oxidation in our films. Film thickness reduction upon illumination was measured with a diamond stylus profilometer. Trends in PC exhibit a bell shaped curve with a rather large effect ( 25%) localized in the Intermediate phase, 20 %< x < 25% of the present binary bulk glass system. The columns observed in SEM investigations reveal a high aspect ratio with length and lateral width of 1.5 µm and about 50 nm diameter. The blue shift (up to 0.38 eV) in bandgap of oblique films (α = 80°) in relation to normally deposited (α = 0°) films, we trace to confinement of electron –hole pairs produced upon optical absorption in nano-filaments. We view the columns formed in films to be composites or bundles of *nanometric sized atomic filaments,* and trace the giant PC effect in the IP to the stress-free character of *filaments of* $Ge_{25}Se_{75}$ *stoichiometry* that undergo facile photomelting resulting in their collapse , and the consequent densification of films. The underlying photostructural effect involves a rapid switching of Ge-Se and Se-Se covalent bonds mediated by a recombination of geminate excitons following absorption of pair-producing radiation. Photomelting of corresponding bulk compositions that belong to the IP has been demonstrated separately [20]. Raman scattering suggests that Ge-rich films ( 25% < x < 33.3%) intrinsically segregate on a nanoscale into $Ge_{25}Se_{75}$ based nanometric composite filaments and a compensating Ge-rich phase ( approx. $Ge_{40}Se_{60}$ stoichiometry) that progressively fills the inter-columnar free space. And the steady loss of PC with increasing x is traced to the growth of the Ge-rich phase, which is stressed and thus photo-inactive. In contrast, Raman scattering data show Se-rich films ( x < 20%) to be homogeneous, and we identify the loss of PC to the accumulation of network stress in such Se-rich columns, which renders them progressively photo-inactive. Evidence of accumulation of network stress in Se-rich bulk glasses was demonstrated in pressure Raman experiments earlier. In summary, the comprehensive set of measurements on a Ge-Se thin films provide the microscopic origin of the giant PC effects in thin-films of the chalcogenides. These findings provide a good basis to predict trends in PC on other chalcogenide glass systems where the IP has been established [24].

**Acknowledgements**

We acknowledge discussions with Professor Bernard Goodman and Professor Peter Smirniotis during the course of this work. This work was supported by the NSF grant DMR 04-56472.

## Tables

Table 1: $E_T$ and ($E_{53}$) of $GeSe_2$ thin films.

## Figure captions

**Fig. 1**. Experimental setup for photo-illumination of $Ge_xSe_{1-x}$ thin films. Radiation from a Hg-Xe arc lamp is condensed and collimated to irradiate sample mounted in a cold stage which is water cooled and the sample chamber flushed with $N_2$ gas.

**Fig. 2**. Thickness changes of $Ge_xSe_{1-x}$ thin films after photo- contraction (●) showing a bell shaped curve with a broad maximum near x ~ 23%. The non-reversing enthalpy at $T_g$

of corresponding bulk glasses ($\Delta H_{nr}$ ) reveals a global minimum in the 20% < x < 25% range identified as the Intermediate phase. See text for details.

**Fig. 3**. SEM images of (top) normally ($\alpha = 0^o$) and (bottom) obliquely ($\alpha = 80^o$) deposited amorphous $Ge_xSe_{1-x}$ thin films (x = 33.3%). Columnar growth in the obliquely deposited film is observed as indicated by the arrows.

**Fig. 4**. Raman scattering of bulk $GeSe_2$ bulk glass, and amorphous films at indicated angle (α) of deposition. Note that the normally deposited (α= 0) film shows a Raman spectrum similar to that of the bulk glass used as the evaporation charge. Raman lineshapes change systematically as α increases and permit network structure to be decoded. See text.

**Fig. 5**. Raman scattering of bulk $Ge_{30}Se_{70}$ glass, normally α= 0) and obliquely deposited films at α = 45°, $60^o$ and $80^o$ showing a pattern similar to the one seen for $GeSe_2$ films in Figure 4.

**Fig. 6**. Raman scattering of bulk $Ge_{23}Se_{77}$ glass, normally deposited (α= 0), and obliquely deposited films at α = $60^o$ and $80^o$ showing no change as a function of α.

**Fig. 7**. Results of Raman lineshape analysis yielding variations in (a) $A_1$ mode frequency; (b) $A^c_1$ mode frequency and (c) Raman intensity ratio of $A^c_1$ / $A_1$ as function of obliqueness angle. Results for amorphous $Ge_xSe_{1-x}$ thin-films at indicated compositions 'x' are plotted.

**Fig. 8**. Results of Raman lineshape analysis on bulk $Ge_xSe_{1-x}$ glasses, showing plots of (a) $A_1$ mode frequency; (b) $A^c_1$ mode frequency (c) Raman intensity ratio of $A^c_1$ / $A_1$ (d) average frequency of $Se_n$ chain mode and (d) Intensity ration of $Se_n$ chain mode to $A_1$ mode as function of glass composition x. These parameters help in fixing the stoichiometry of the backbone in corresponding films.

**Fig. 9**. Micro-Raman spectra of obliquely deposited ($\alpha$ =80$^o$) $GeSe_2$ films taken before (black) and after ( red or grey) photo-contraction at several depths "d" using a confocal microscope.See text for details.

**Fig. 10**. UV-VIS absorption of normally deposited $Ge_xSe_{1-x}$ films examined as a function of photon energy. Plots of ( a) optical absorbance, (b) optical absorption coefficient, $\mu$ , and (c) of $(\mu h\omega)^{1/2}$ as a function of photon energy appear in different panels. Optical bandgaps of films increase as a function of "x' as illustrated in Figure 11.

**Fig. 11**. Optical band gap of normally deposited $Ge_xSe_{1-x}$ films. (a) The filled circles give the gap $E_{53}$, measured at $\mu$ = 5 x10$^3$ cm$^{-1}$ , while the open circles are data of Street et al. ref 37 taken at $\mu$ = 1 x10$^3$ cm$^{-1}$ . (b) Filled circles give the Tauc edge, $E_T$, of the present films as a function of x. Filled and open squares represents respectively $E_T$ of amorphous x= 33.3% thin-films in their as deposited and after optical annealing taken from the work of Spense and Elliott, ref. 9.

**Fig. 12**. Plots of (a) UV-Vis absorbance and (b) absorption coefficient μ of $Ge_xSe_{1-x}$ thin films at x = 22%, as a function of photon energy for films at indicated obliqueness angle α. Plots of (c) $(\mu h\omega)^{1/2}$ against E, and (d) variations in the optical band gaps $E_{53}$ and $E_T$ of x = 22% films as a function of obliqueness angle.

**Fig. 13**. Plots of (a) UV-Vis absorbance and (b) absorption coefficient μ of $Ge_xSe_{1-x}$ thin films at x = 23%, as a function of photon energy for films at indicated obliqueness angle α. Plots of (c) $(\mu h\omega)^{1/2}$ against E, and (d) variations in the optical band gaps $E_{53}$ and $E_T$ of x = 23% films as a function of obliqueness angle.

**Fig. 14**. (Color on line) Plots of (a) UV-Vis absorbance and (b) absorption coefficient μ of $Ge_xSe_{1-x}$ thin films at x = 30%, as a function of photon energy for films at indicated obliqueness angle α. Plots of (c) $(\mu h\omega)^{1/2}$ against E, and (d) variations in the optical band gaps $E_{53}$ and $E_T$ of x = 30% films as a function of obliqueness angle.

**Fig. 15**. Optical band gap $E_{53}$ of normally and obliquely deposited $Ge_xSe_{1-x}$ thin films as a function of 'x'. Note the much higher value of the gap in obliquely deposited films.

**Fig. 16**. IR absorbance of (a) $GeO_2$ glass and amorphous $Ge_{22}Se_{78}$ films before and after irradiation, and (b) $GeO_2$ glass and amorphous $Ge_{23}Se_{77}$ films before and after irradiation.

In (a) the cold stage was thouroughly flushed with N2 gas while in (b) this was not the case. Note presence of oxidation in the 23% film in (b) but the absence of it in the 22% film in (a).

**Fig. 17**. Simulation of the observed Raman lineshape of an obliquely (α = 80) deposited $GeSe_2$ film in terms of two contributions, one of $Ge_{25}Se_{75}$ bulk glass ( A-nanophase) and the other of $Ge_{36}Se_{64}$ bulk glass ( B-nanophase), revealing nanoscale phase separation of such films. See equation (2) in text.

**Fig. 18**. Schematic representation of nanoscale phase separation of obliquely deposited $GeSe_2$ films composed of columns representing the stress-free A-nanophase and the hashed mark region comprising of the stressed B-nanophase. Here α represents the obliqueness angle of deposition, and β the angle made by columns with respect to the normal. Equation 1 in text relates α with β. In obliquely deposited $GeSe_3$ films only the A –nanophase exists, and columns readily photomelt when irradiated with pair-producing radiation.

Table 1: $E_T$ and ($E_{53}$) of $GeSe_2$ thin films

| $GeSe_2$ thin films | band gap (eV) of as deposited films | band gap (eV) after illumination |
|---|---|---|
| $0^o$ (present work, New Delhi ) | 217 (233) | |
| $0^o$ (present work, Sofia) | 204 (240) | |
| $0^o$ (Spence & Elliott) | 206 ± 001 | 213±001 |
| $80^o$ (present work, New Delhi) | 205 (268) | 204 (267) |
| $80^o$ (present work, Sofia) | 193 (228) | |
| $80^o$(Spence & Elliott) | 184 ± 004 | 169 ± 004 |

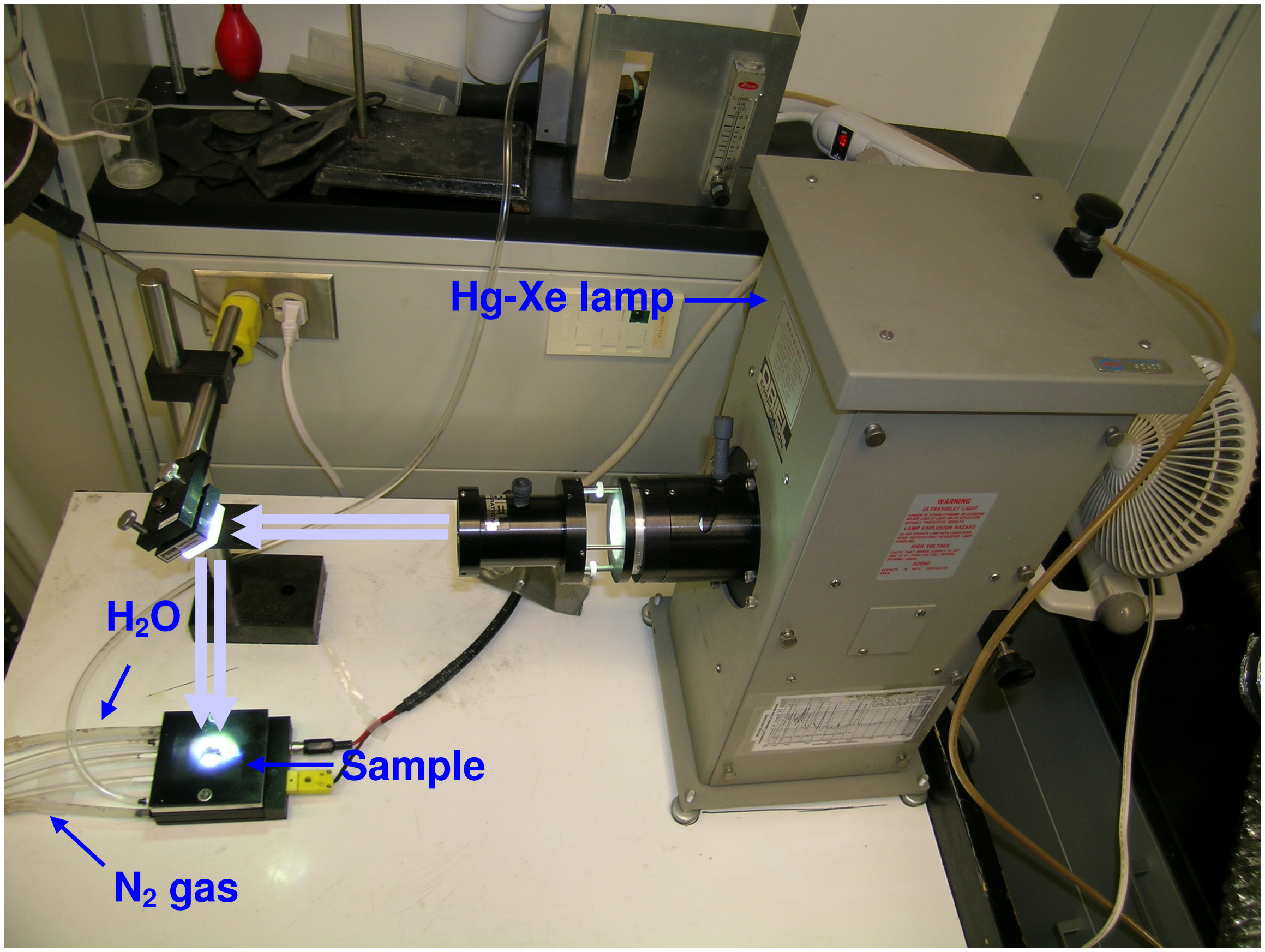

Figure 1

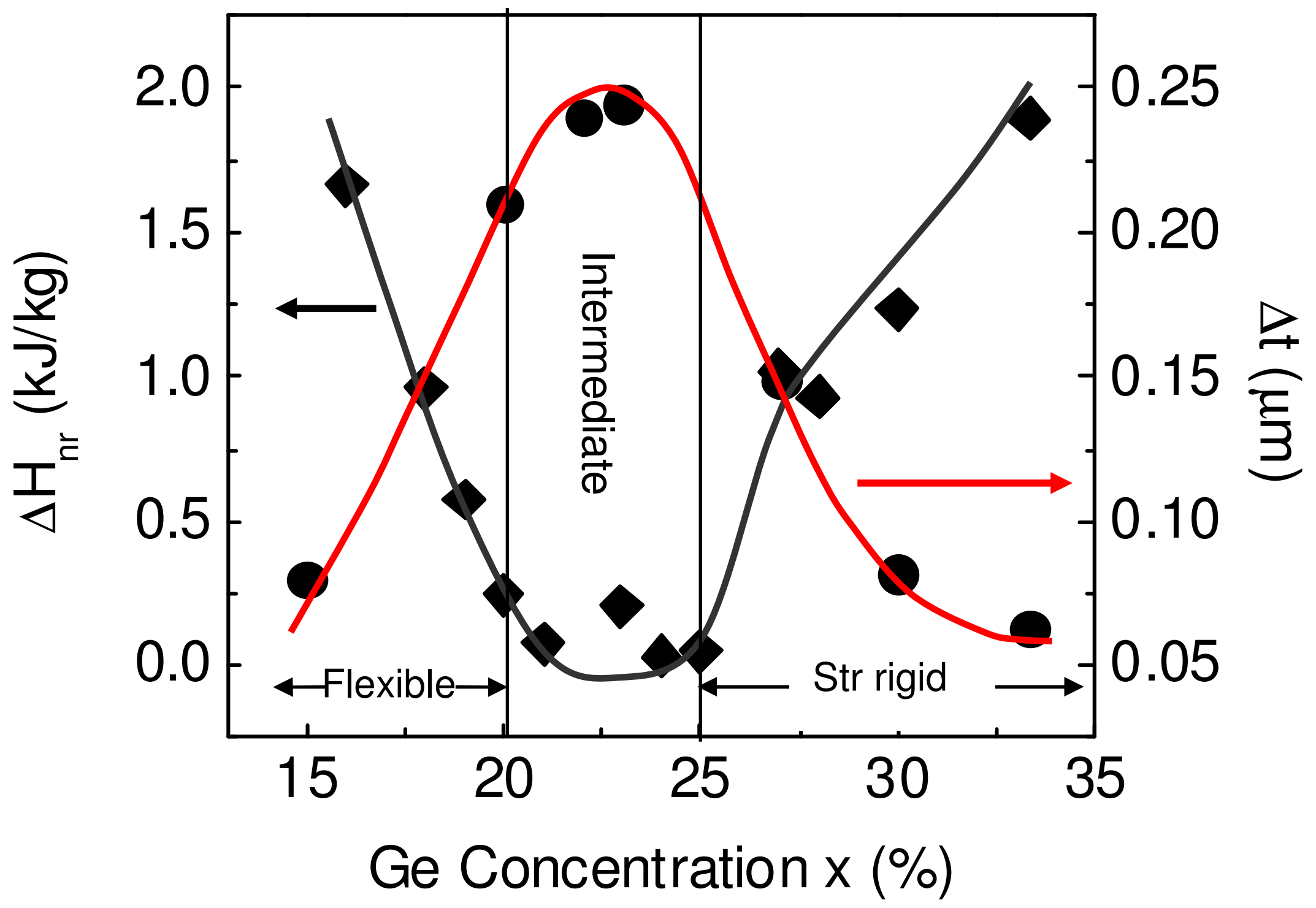

Figure 2

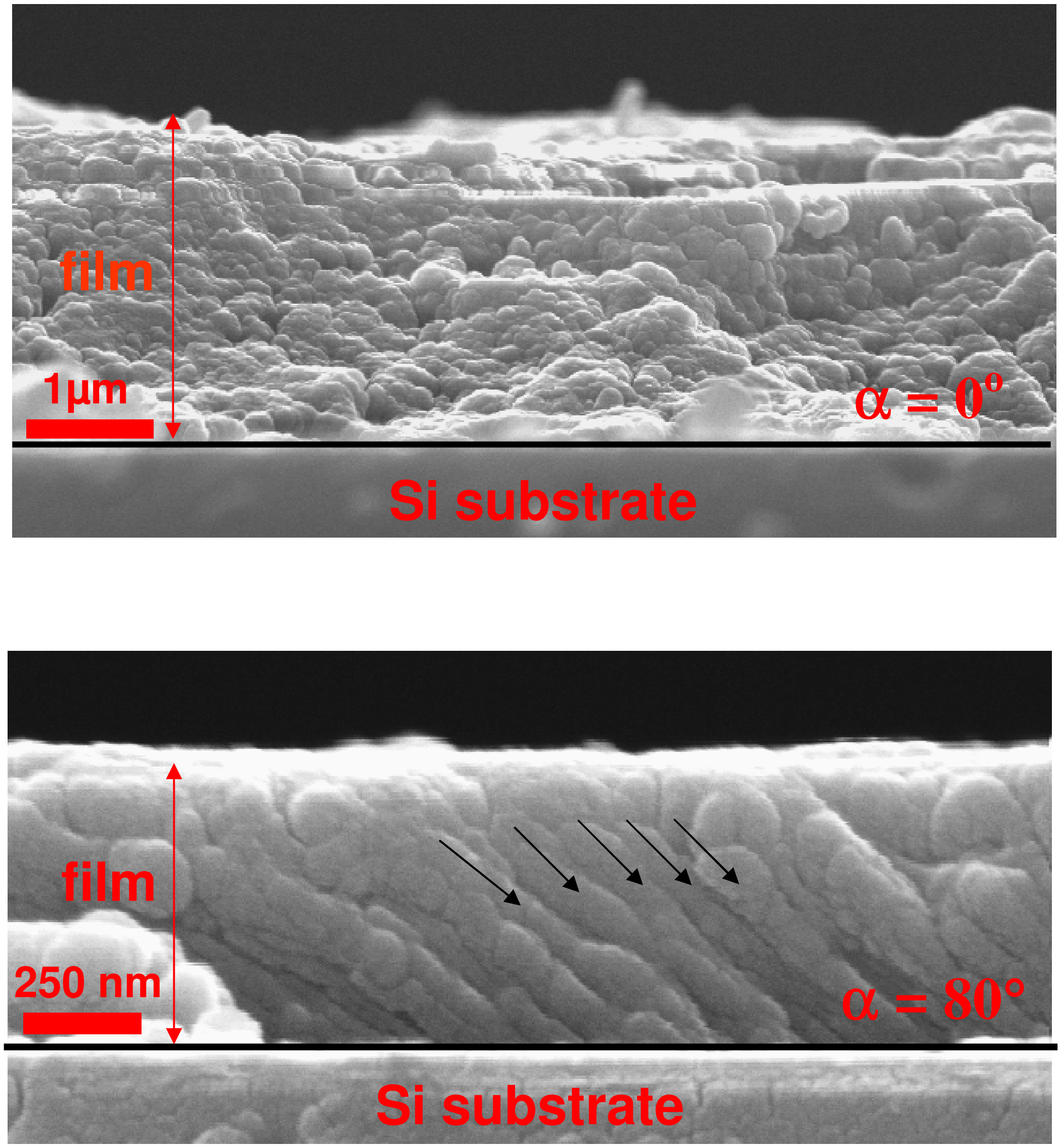

Figure 3

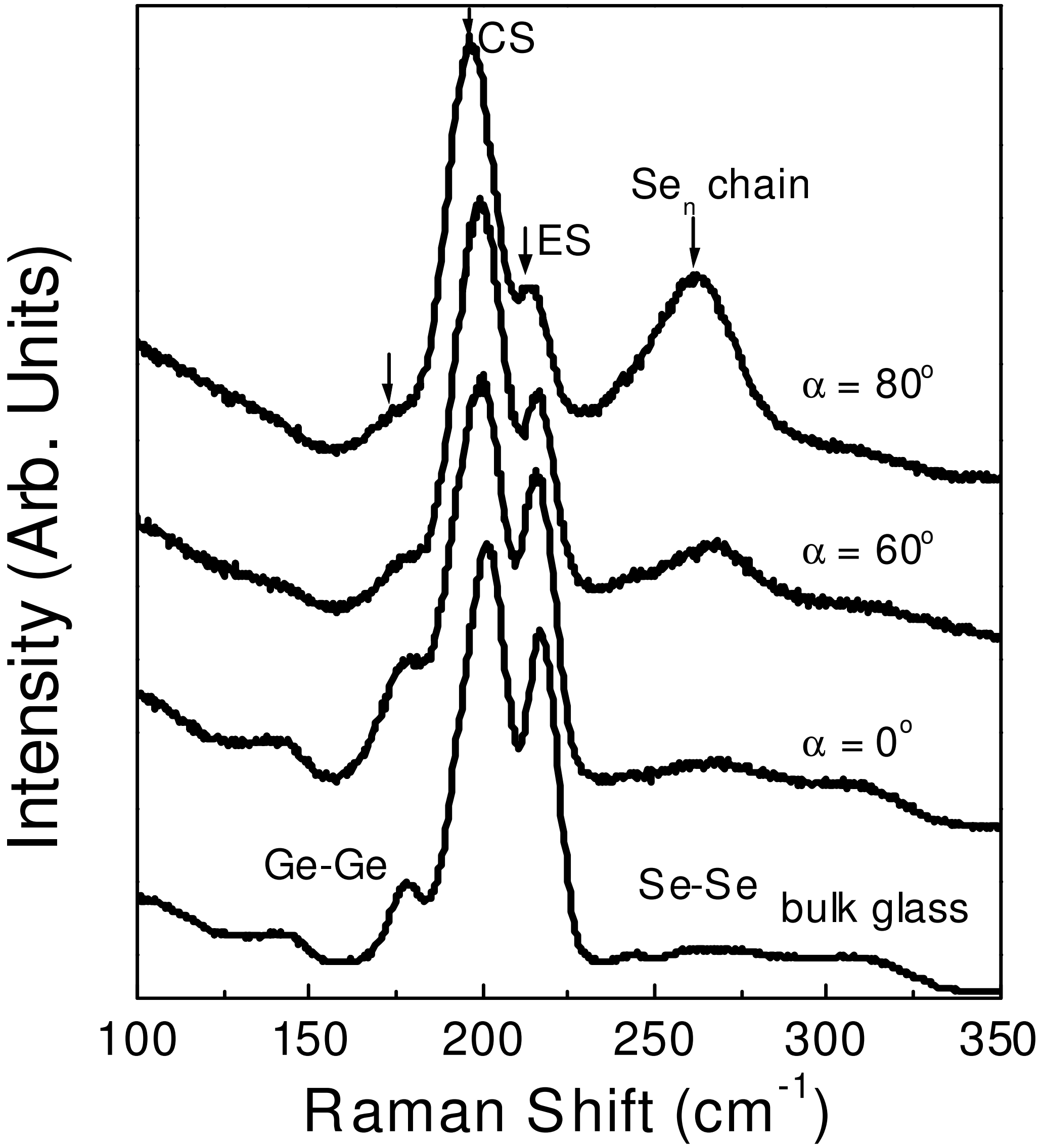

Figure 4

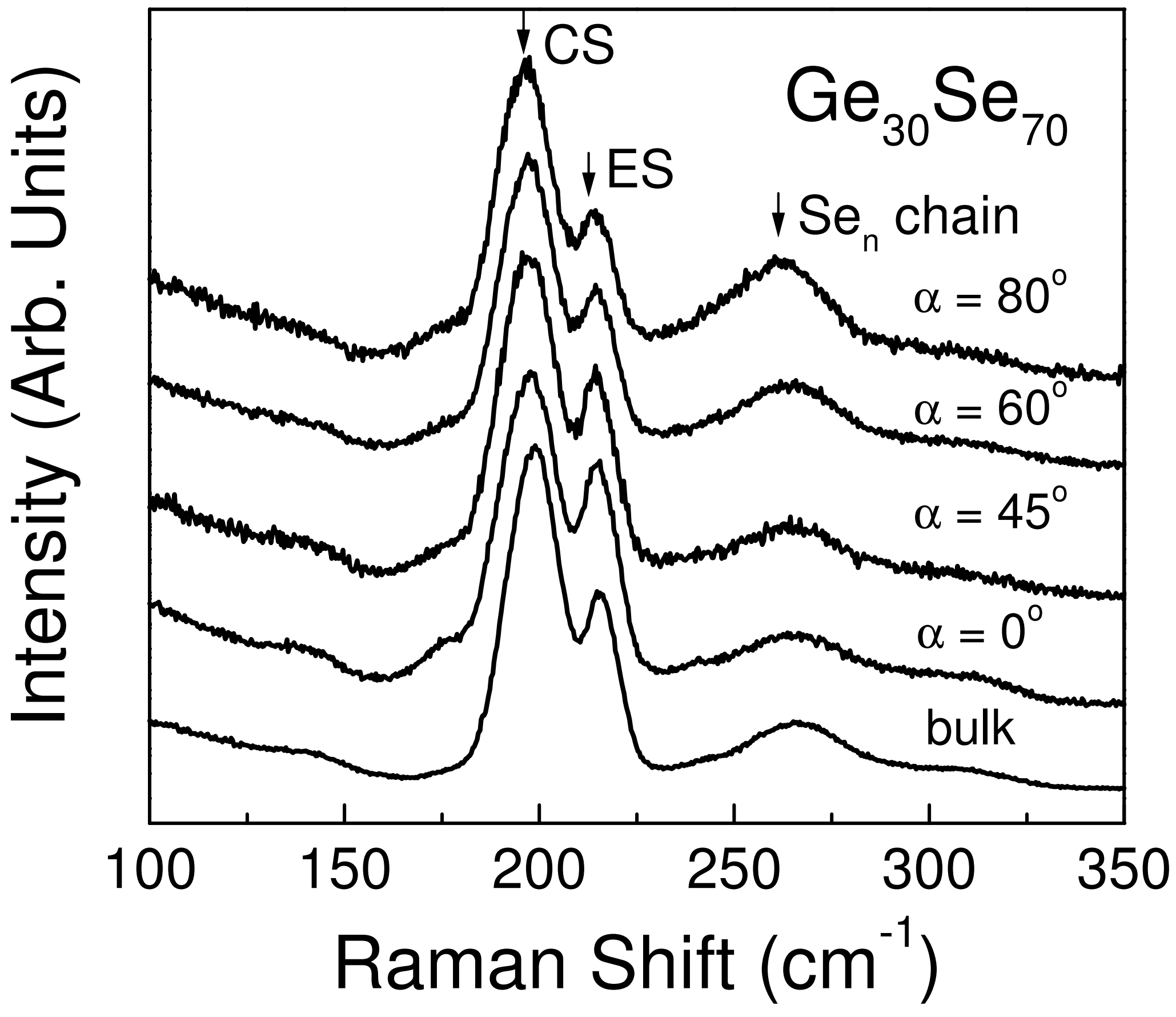

Figure 5

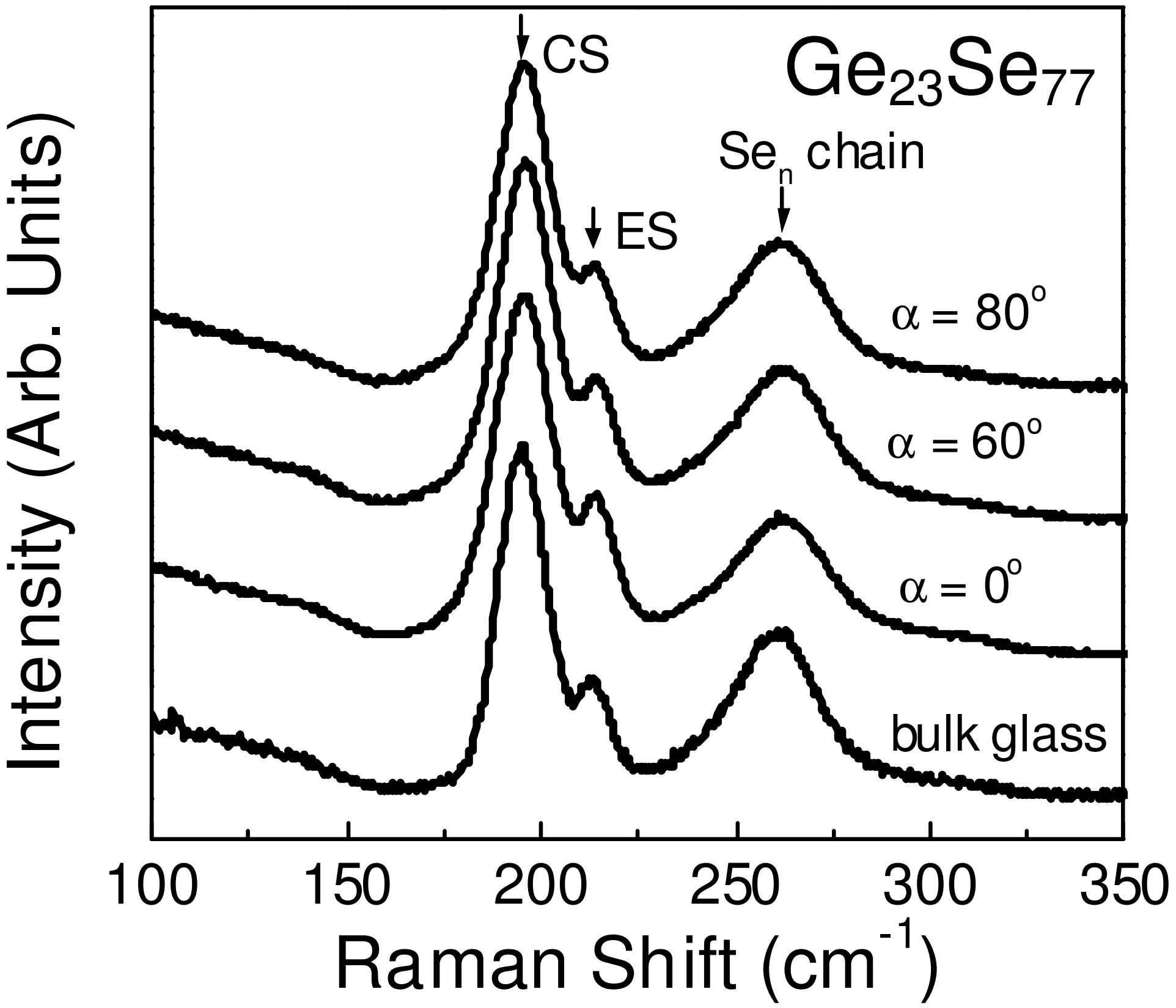

Figure 6

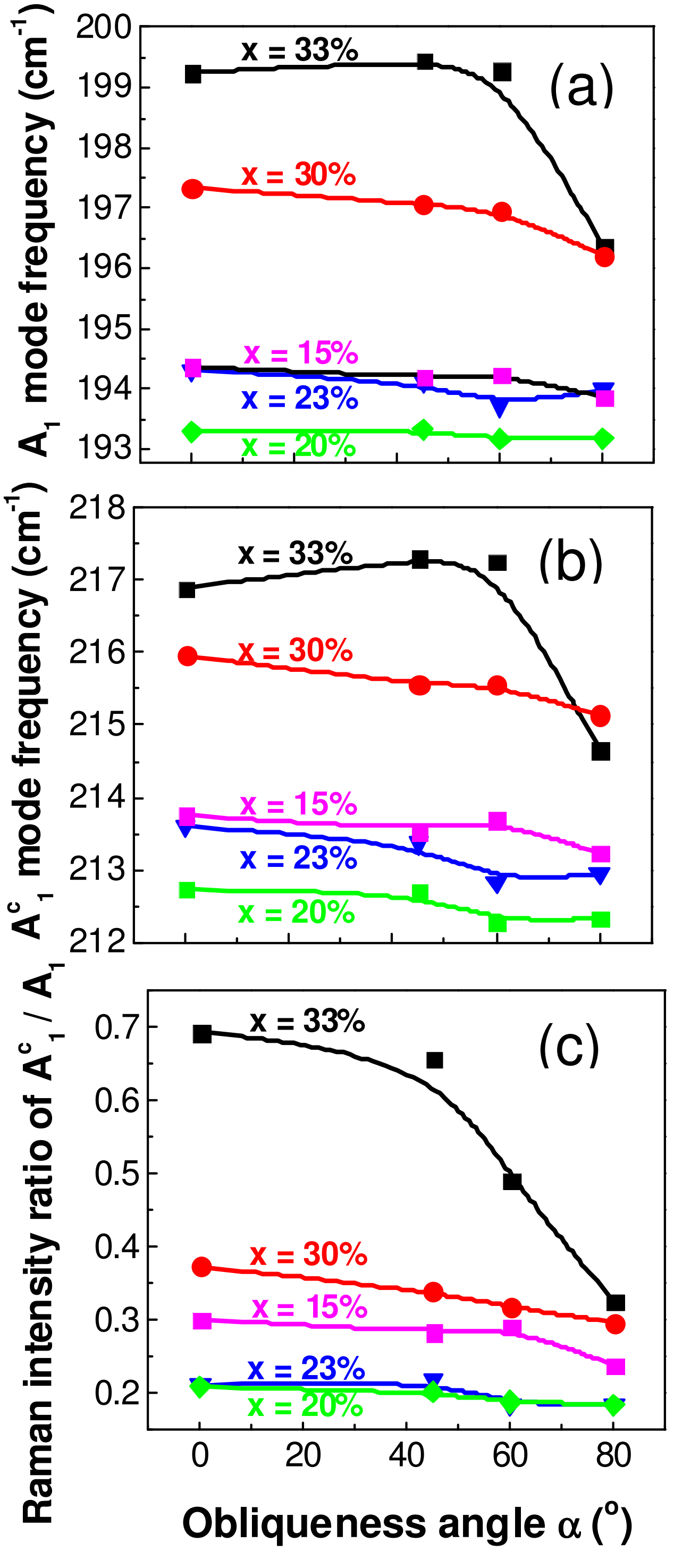

Figure 7

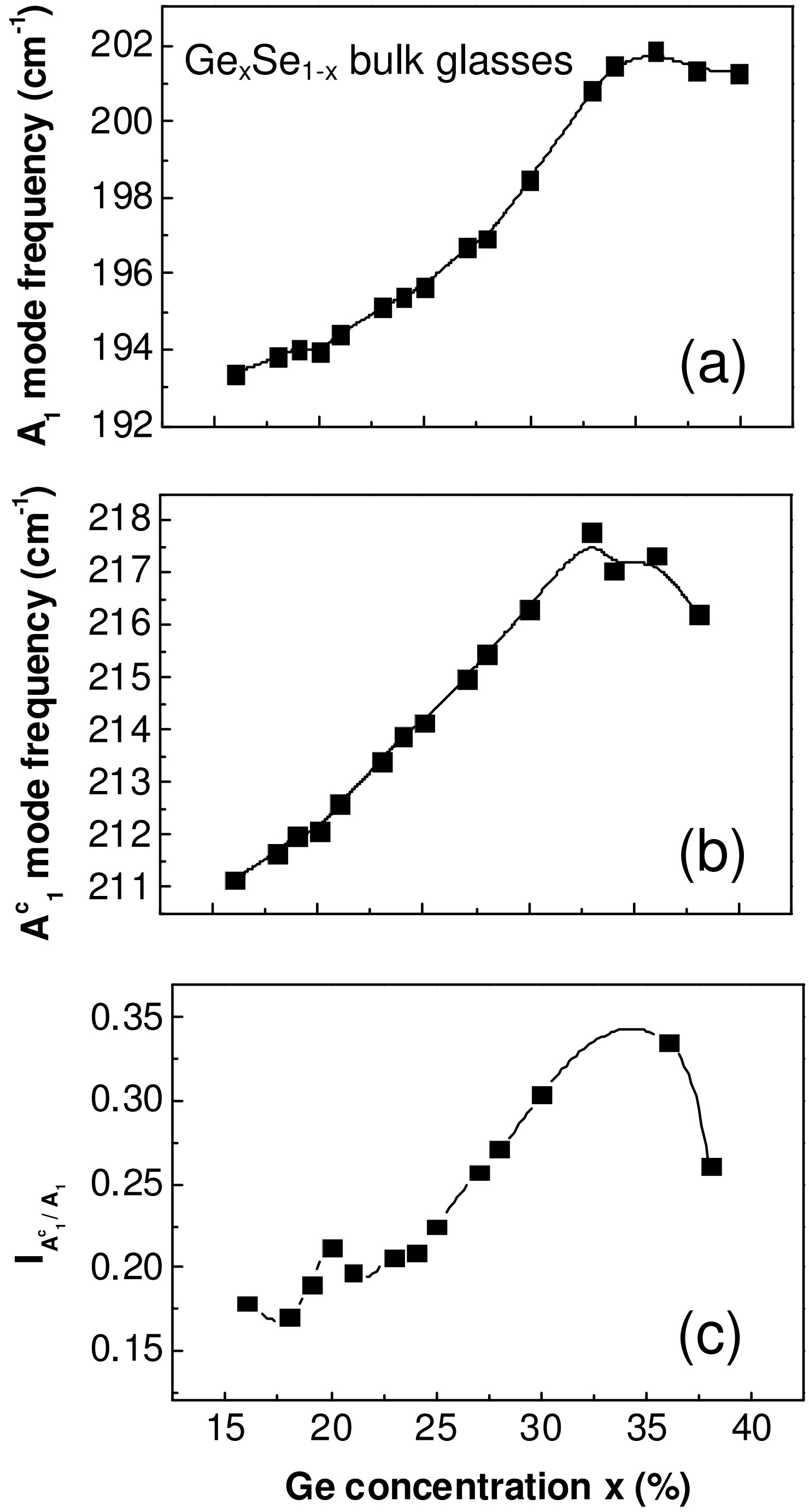

Figure 8

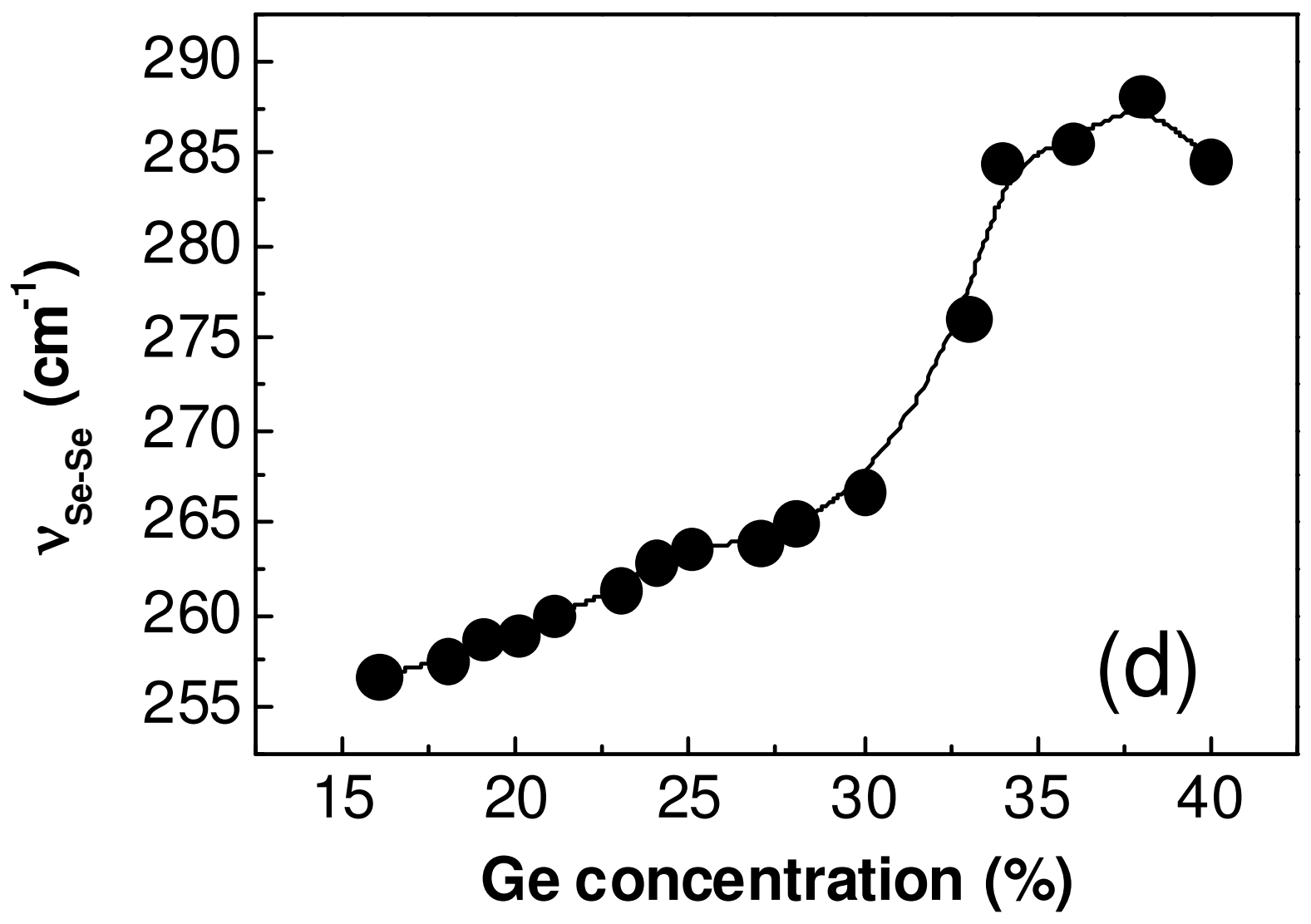

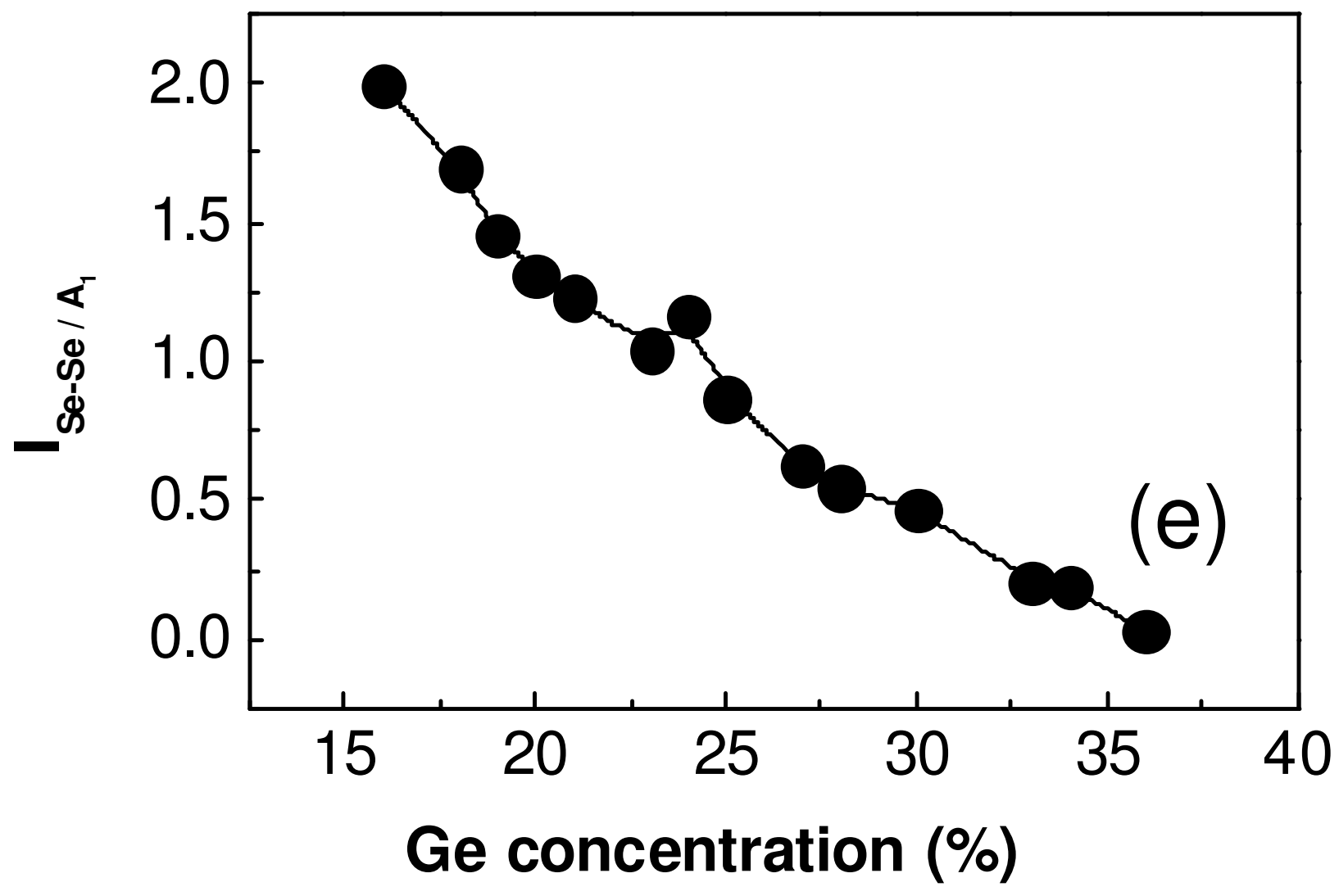

Figure 8 Continued

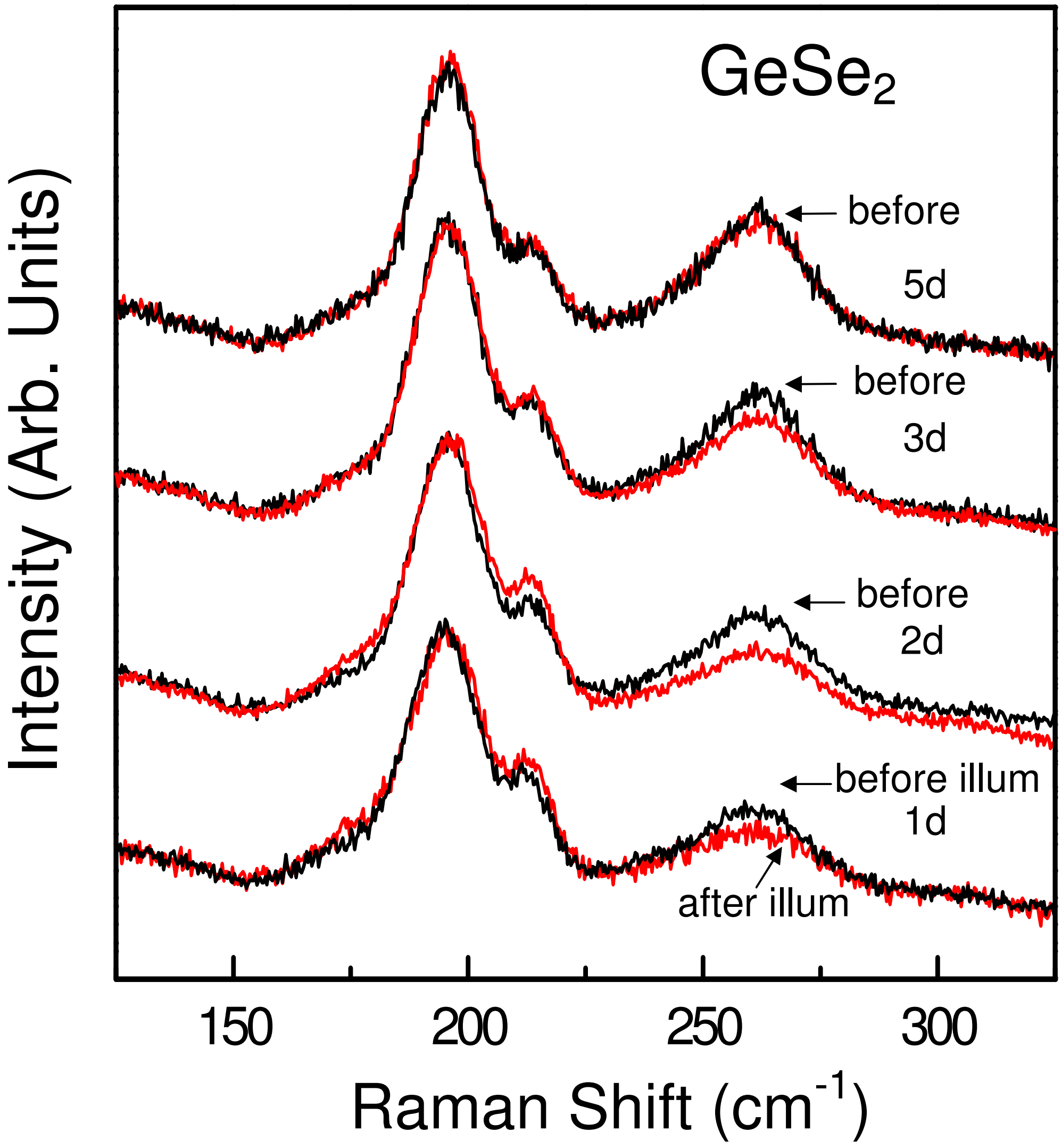

Figure 9

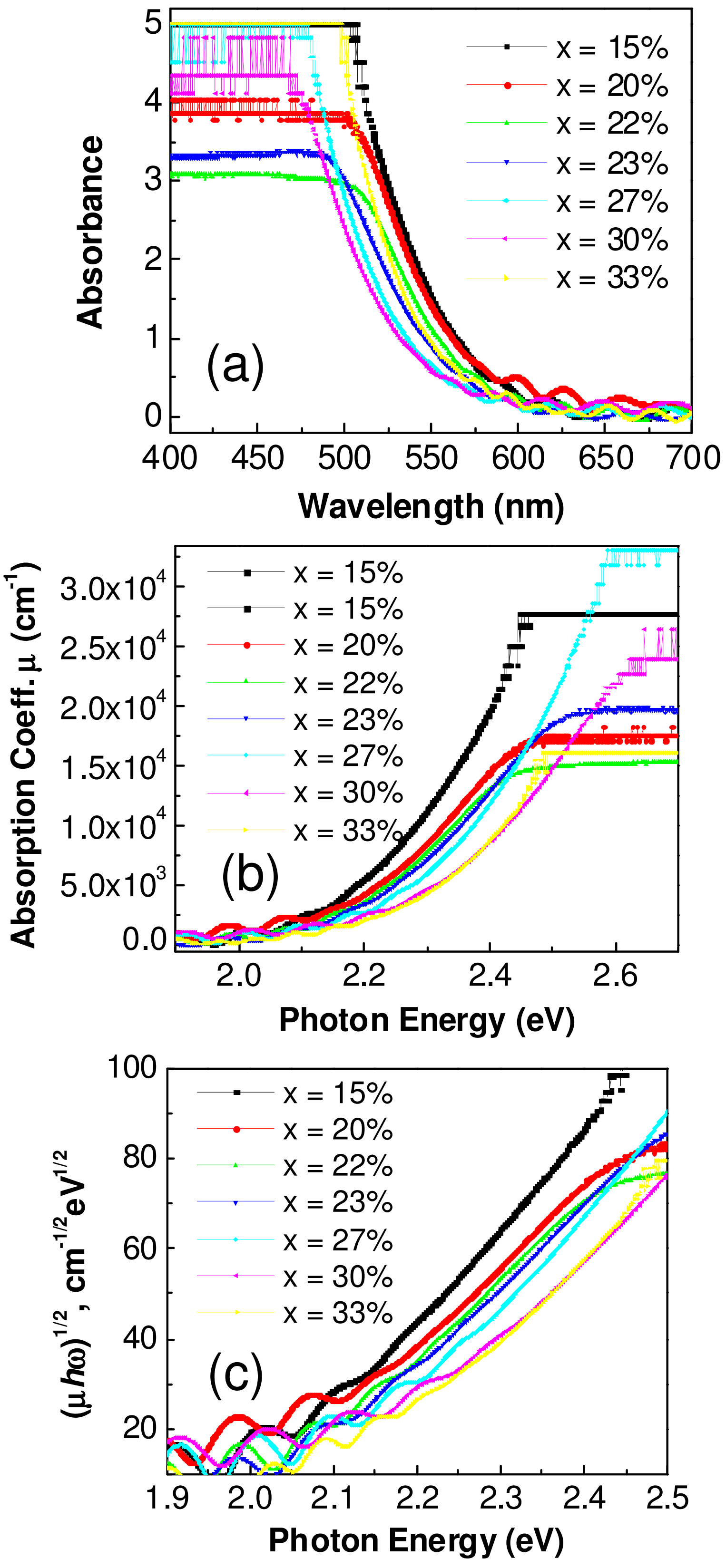

Figure 10

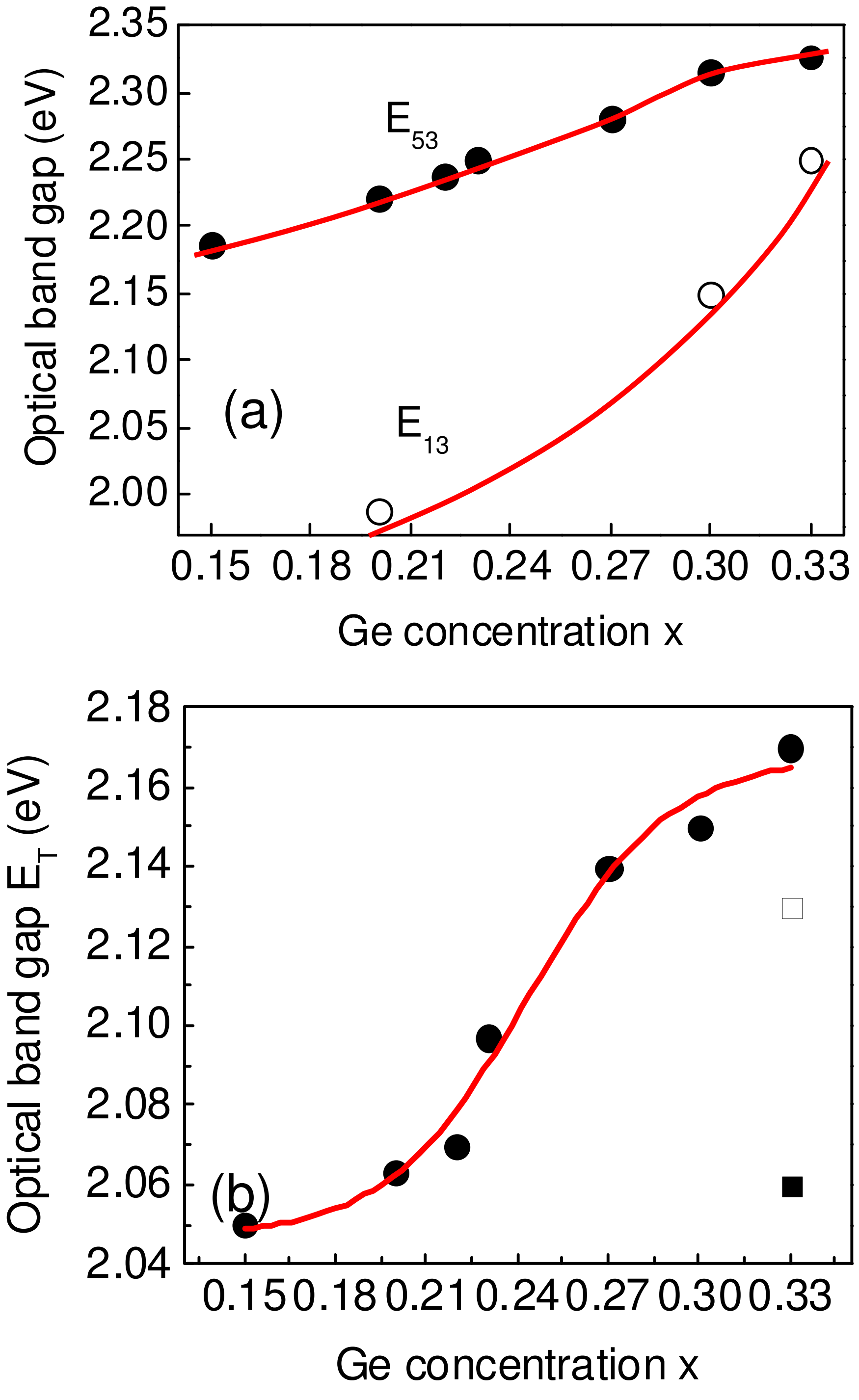

Figure 11

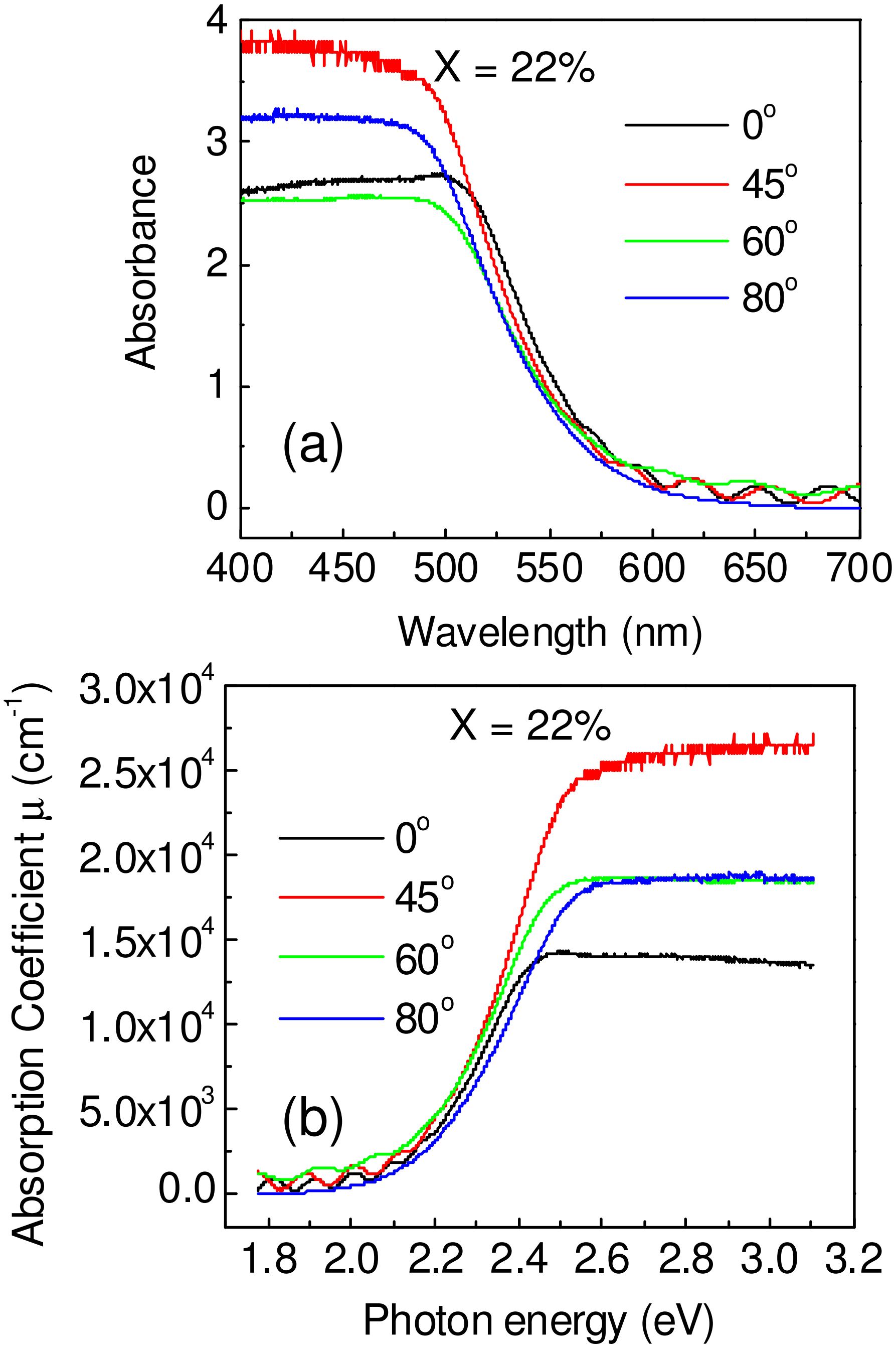

Figure 12

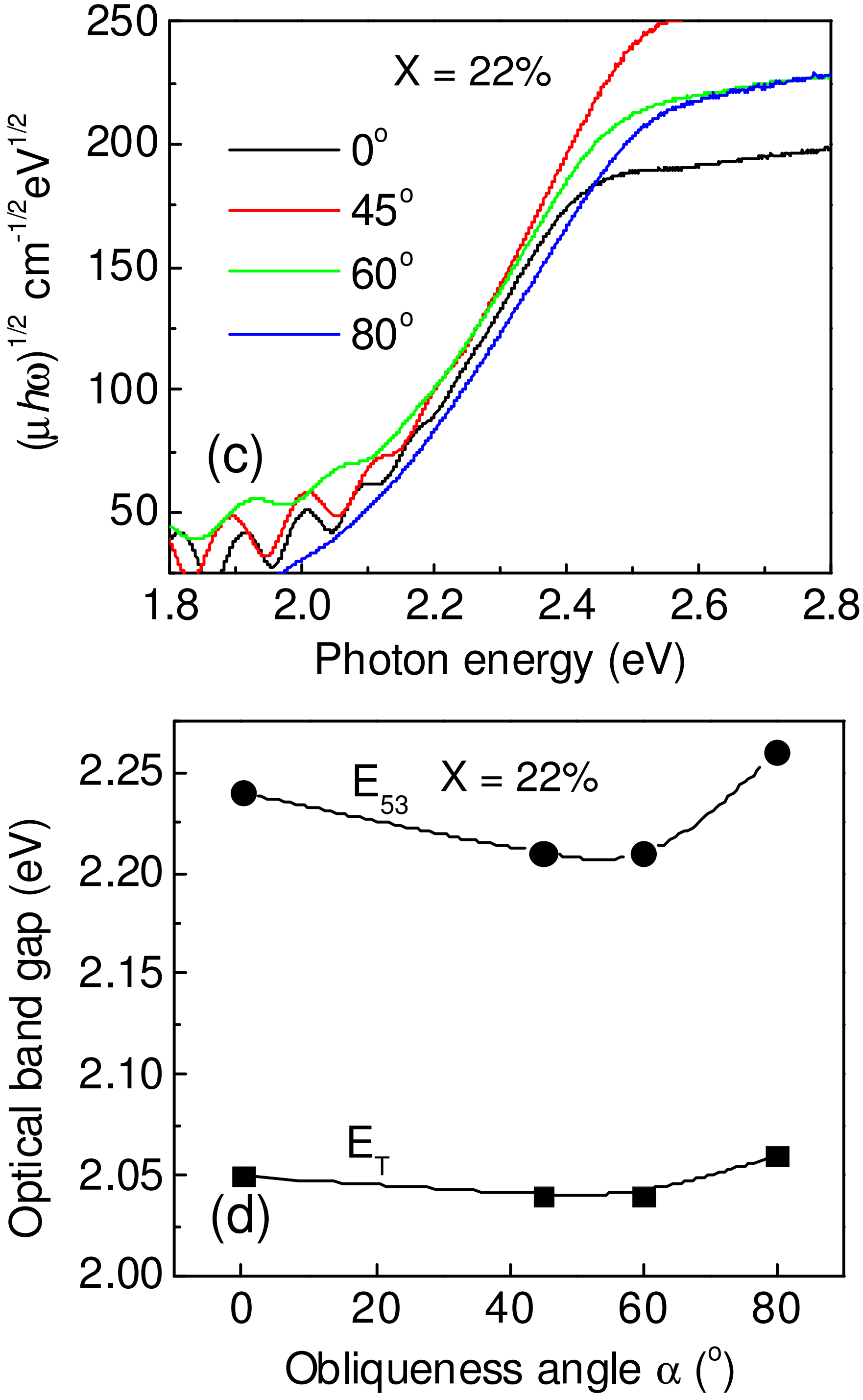

Figure 12 Continued

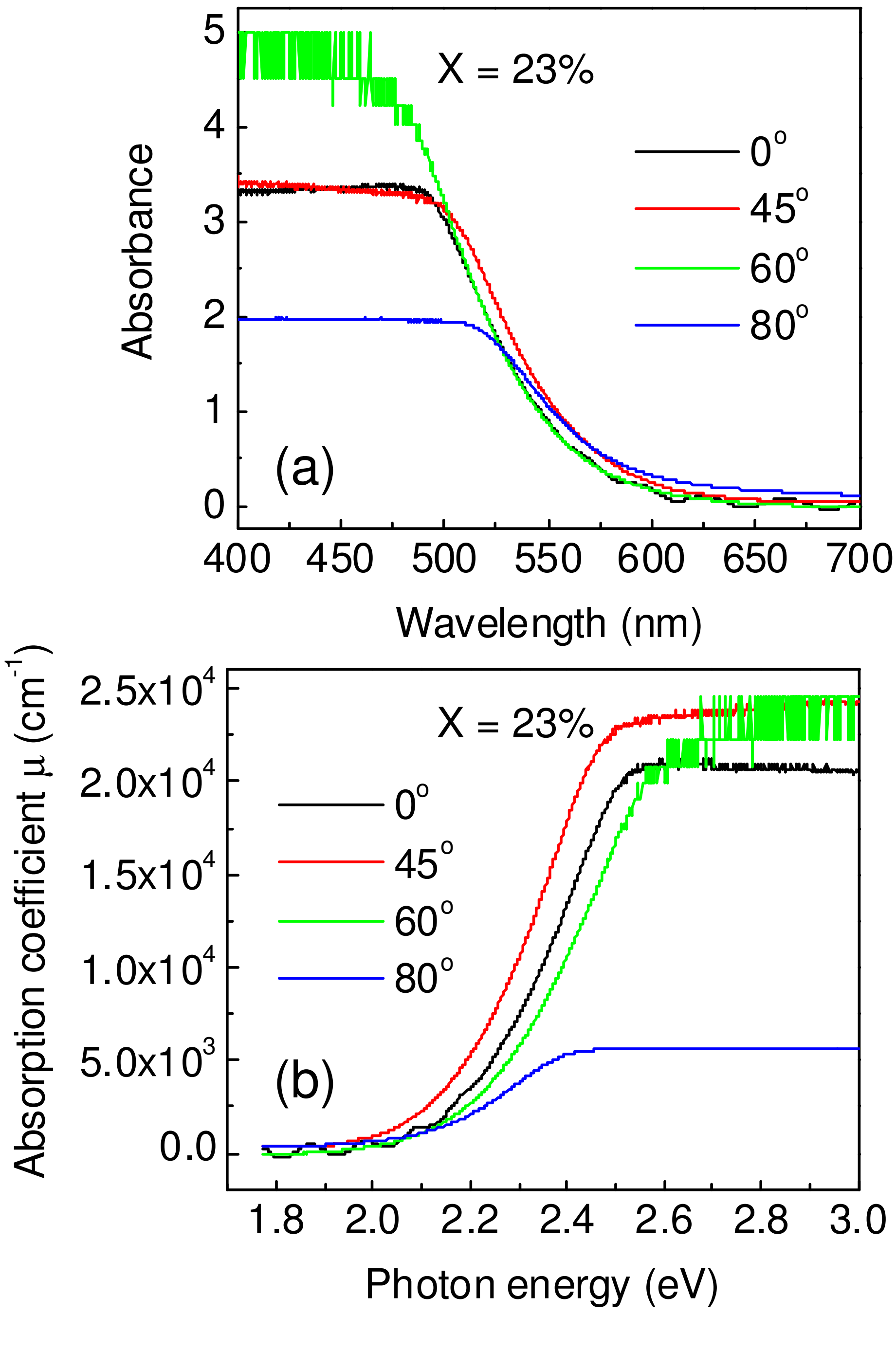

Figure 13

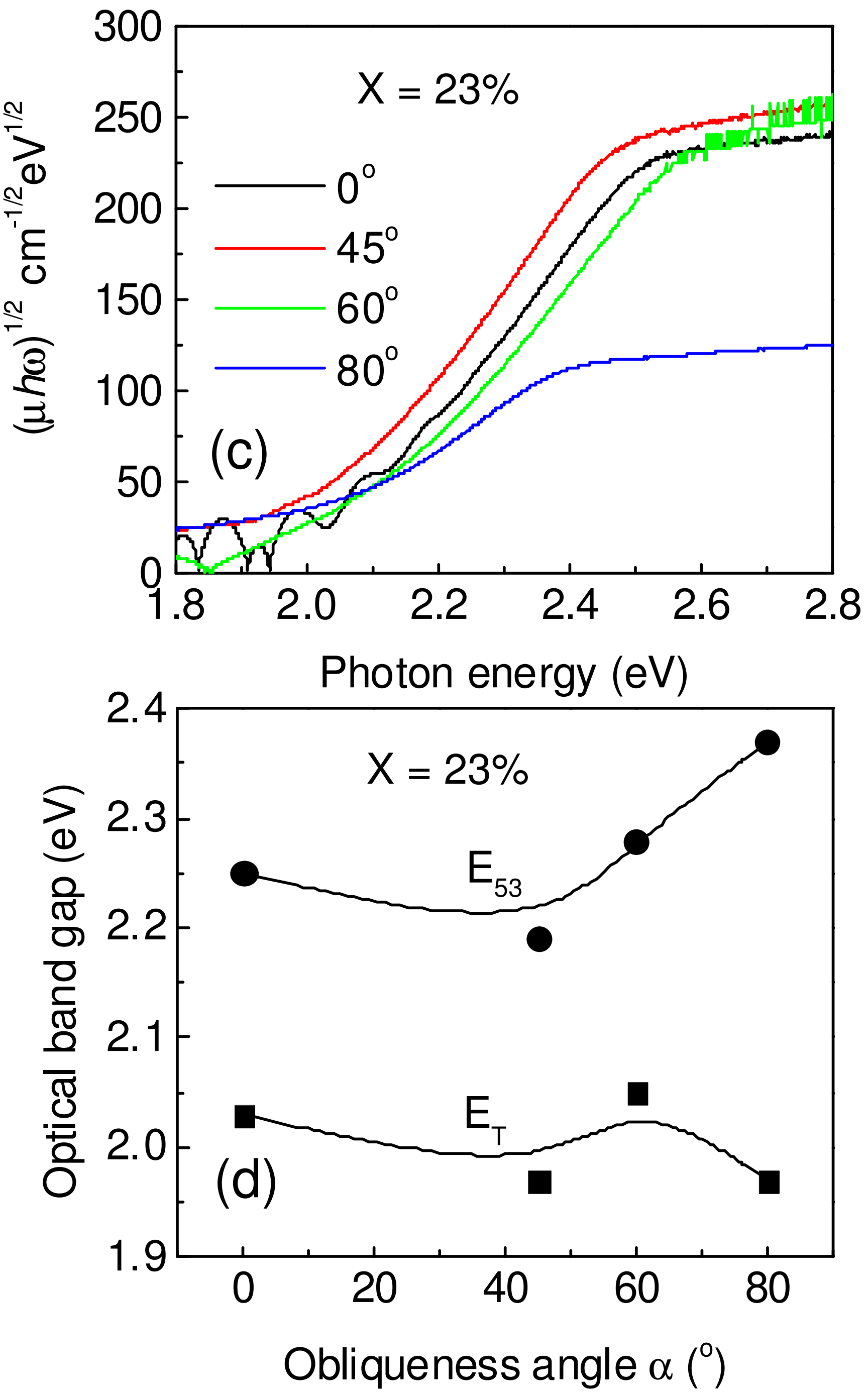

Figure 13 Continued

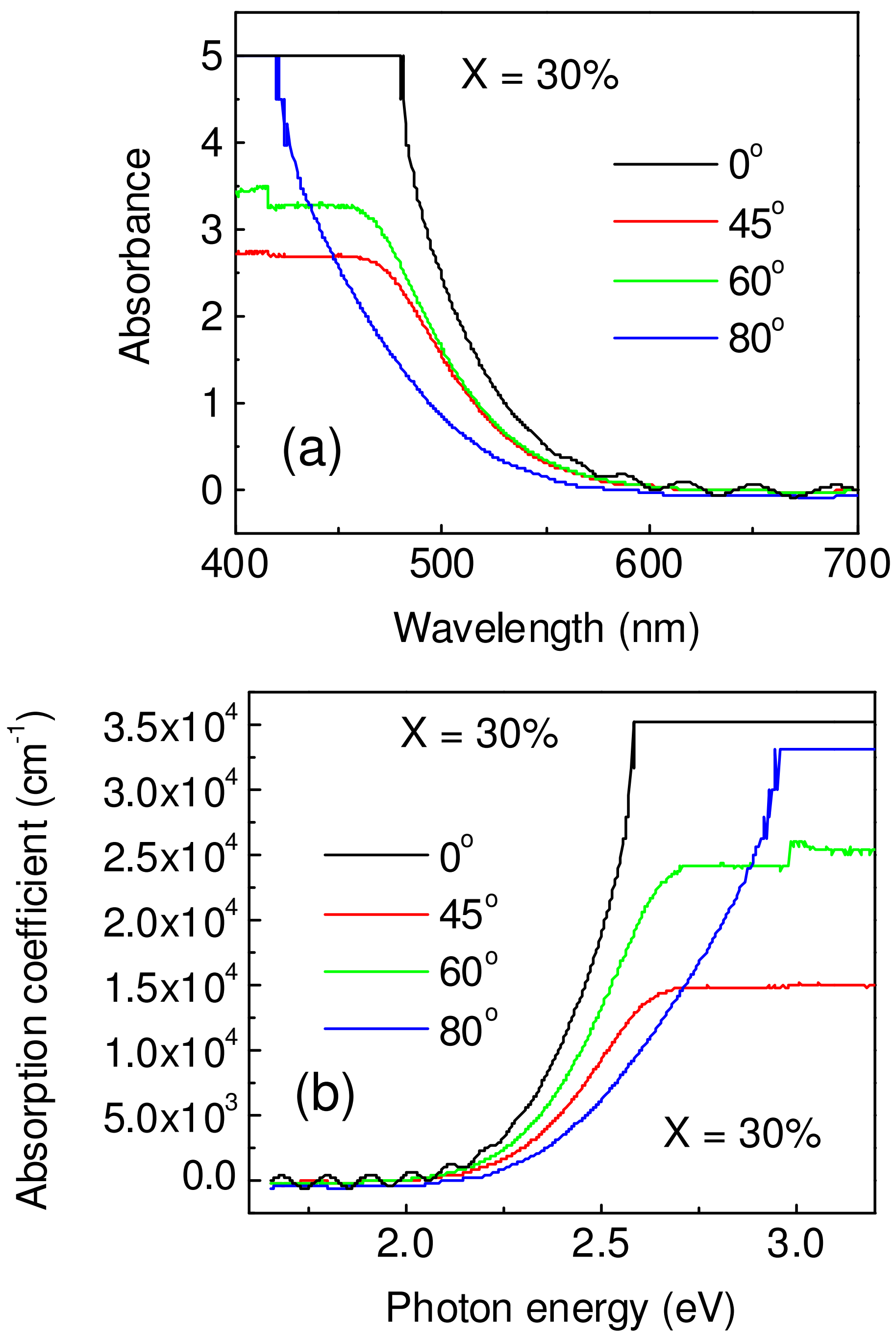

Figure 14

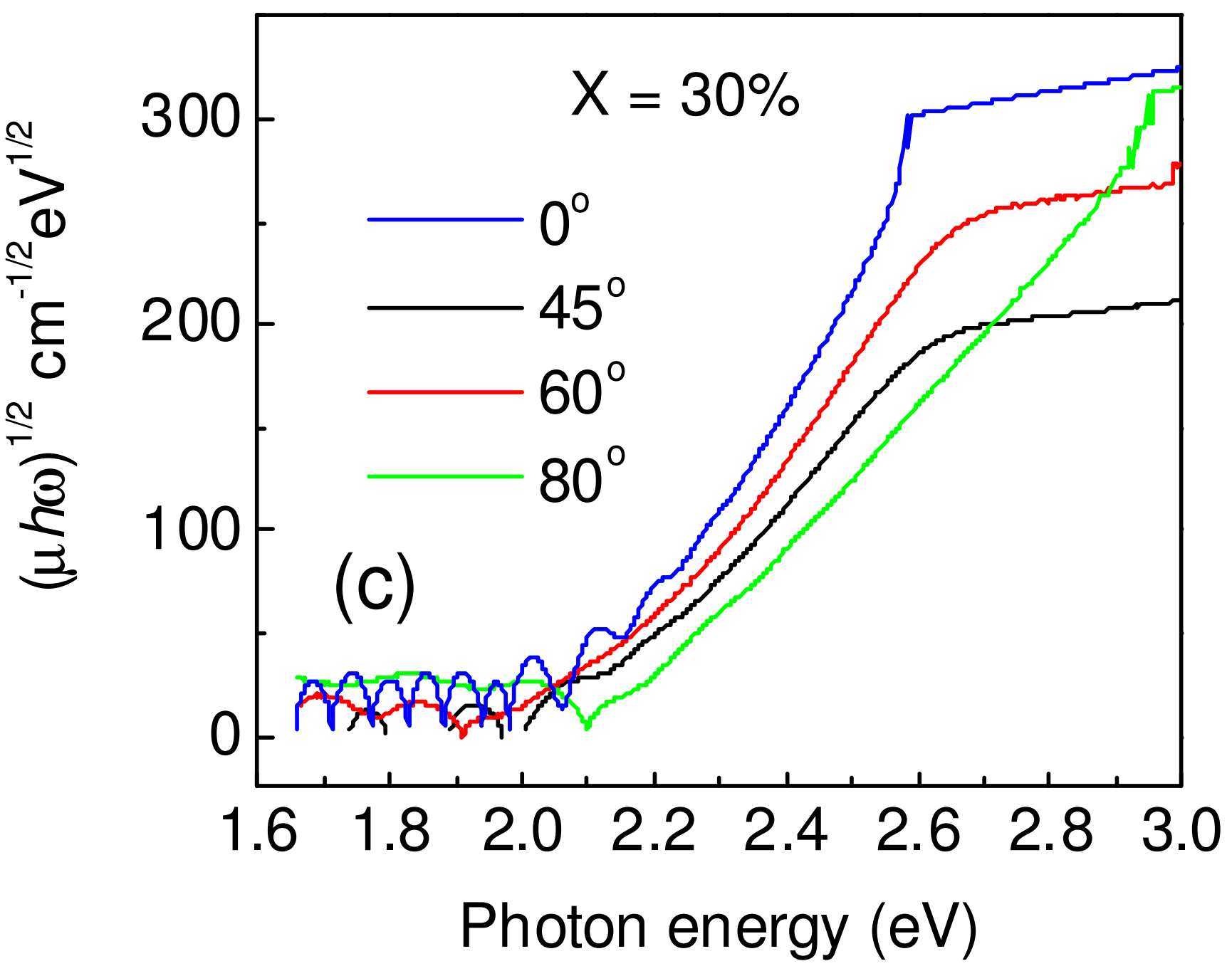

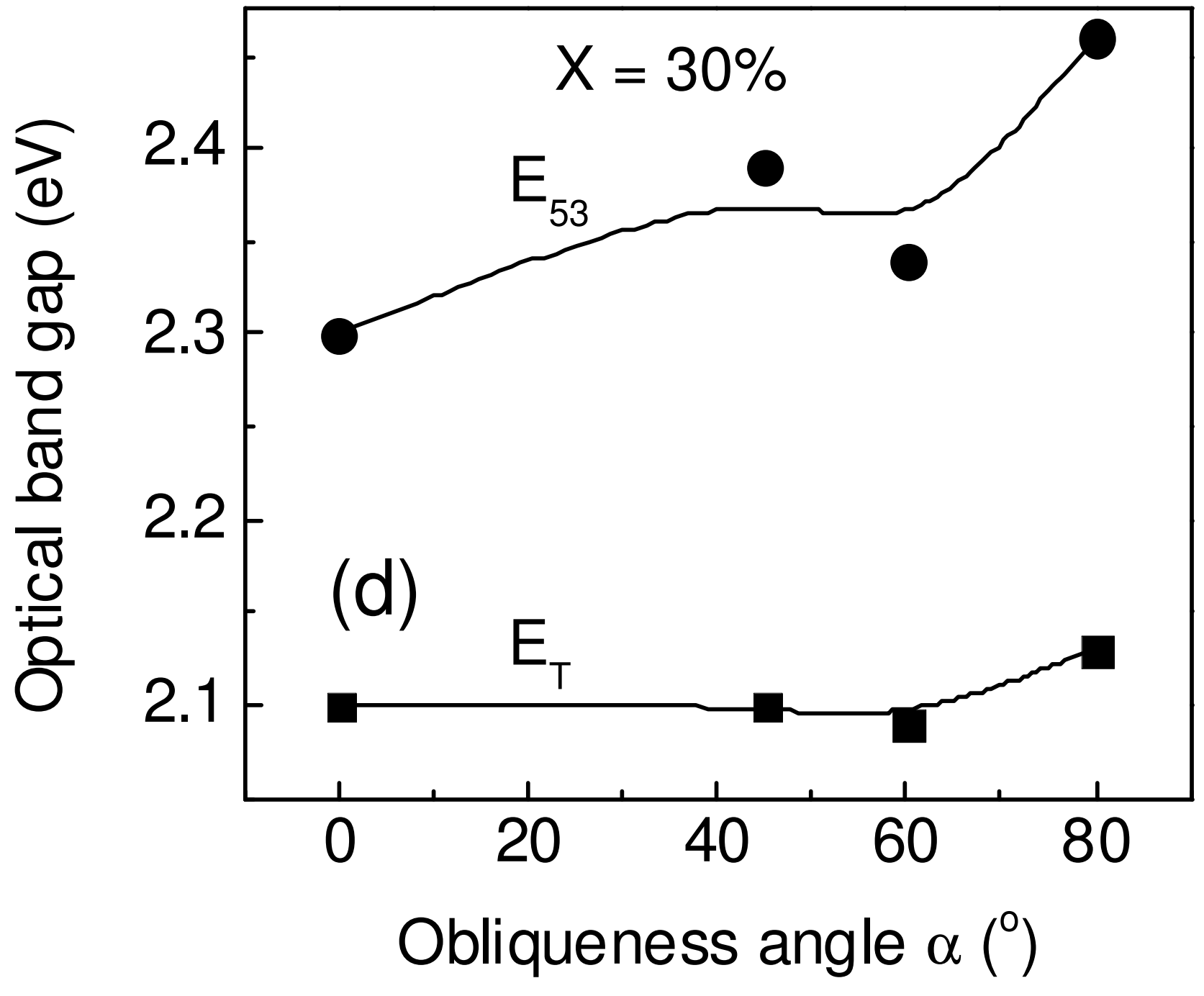

Figure 14 Continued

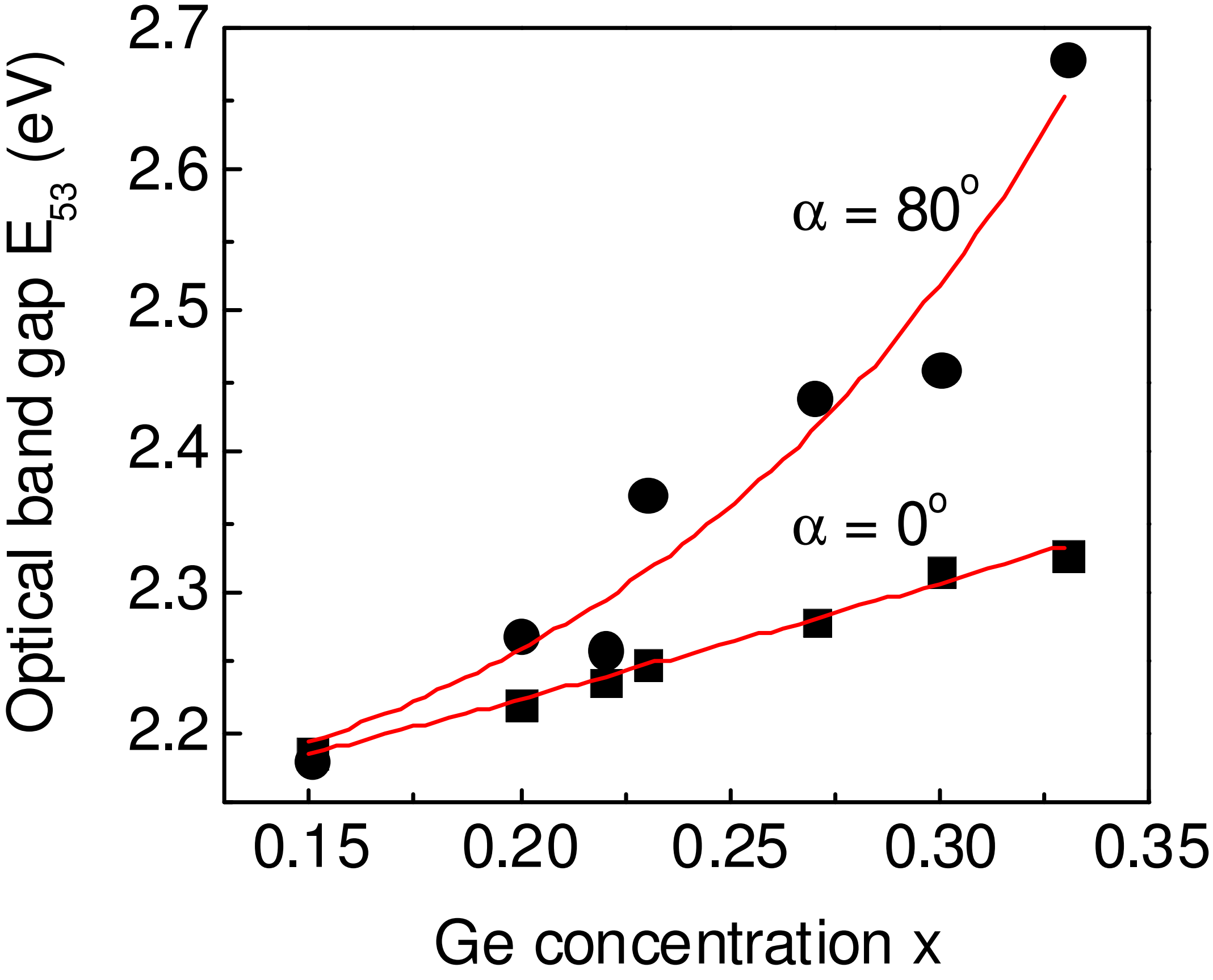

Figure 15

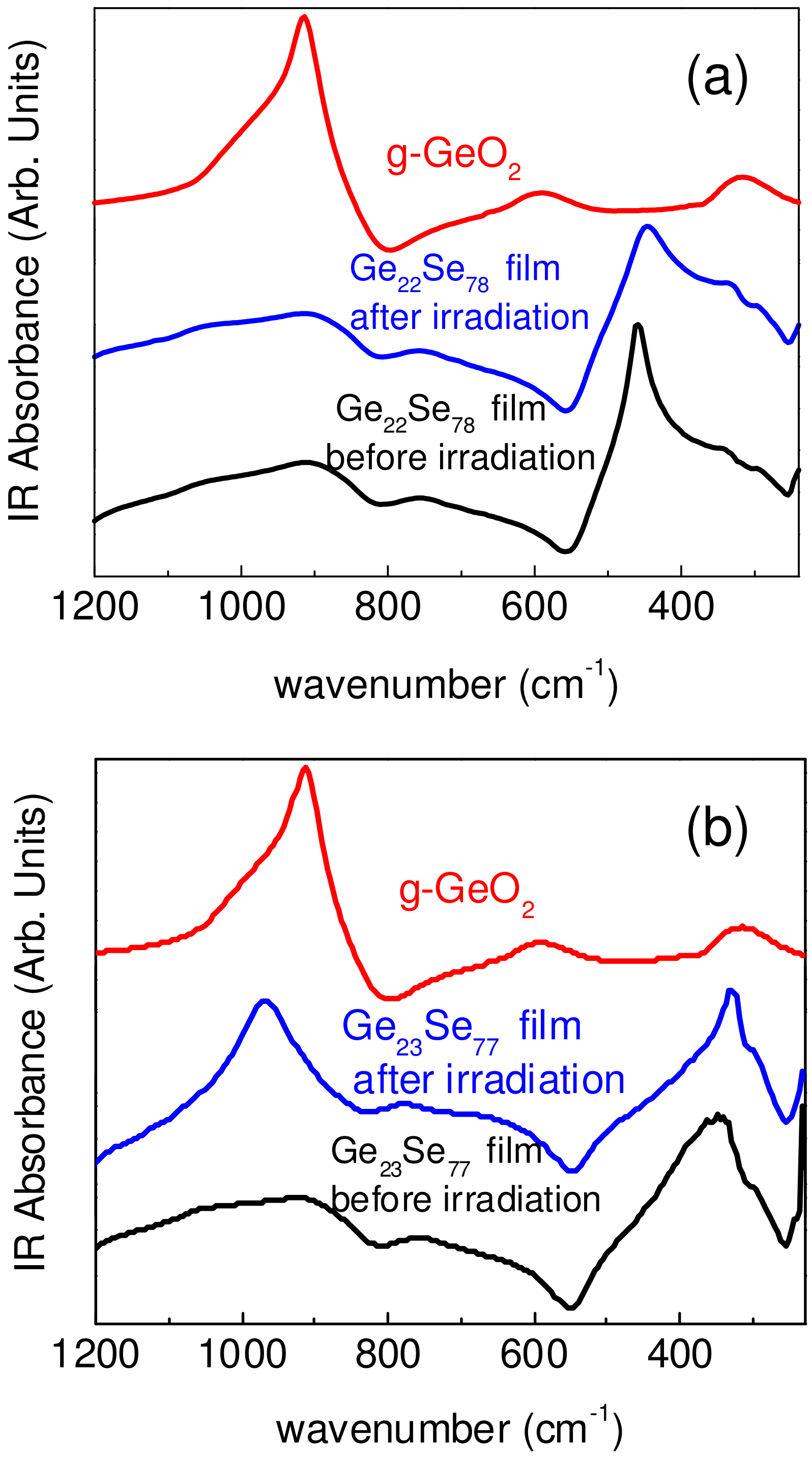

Figure 16

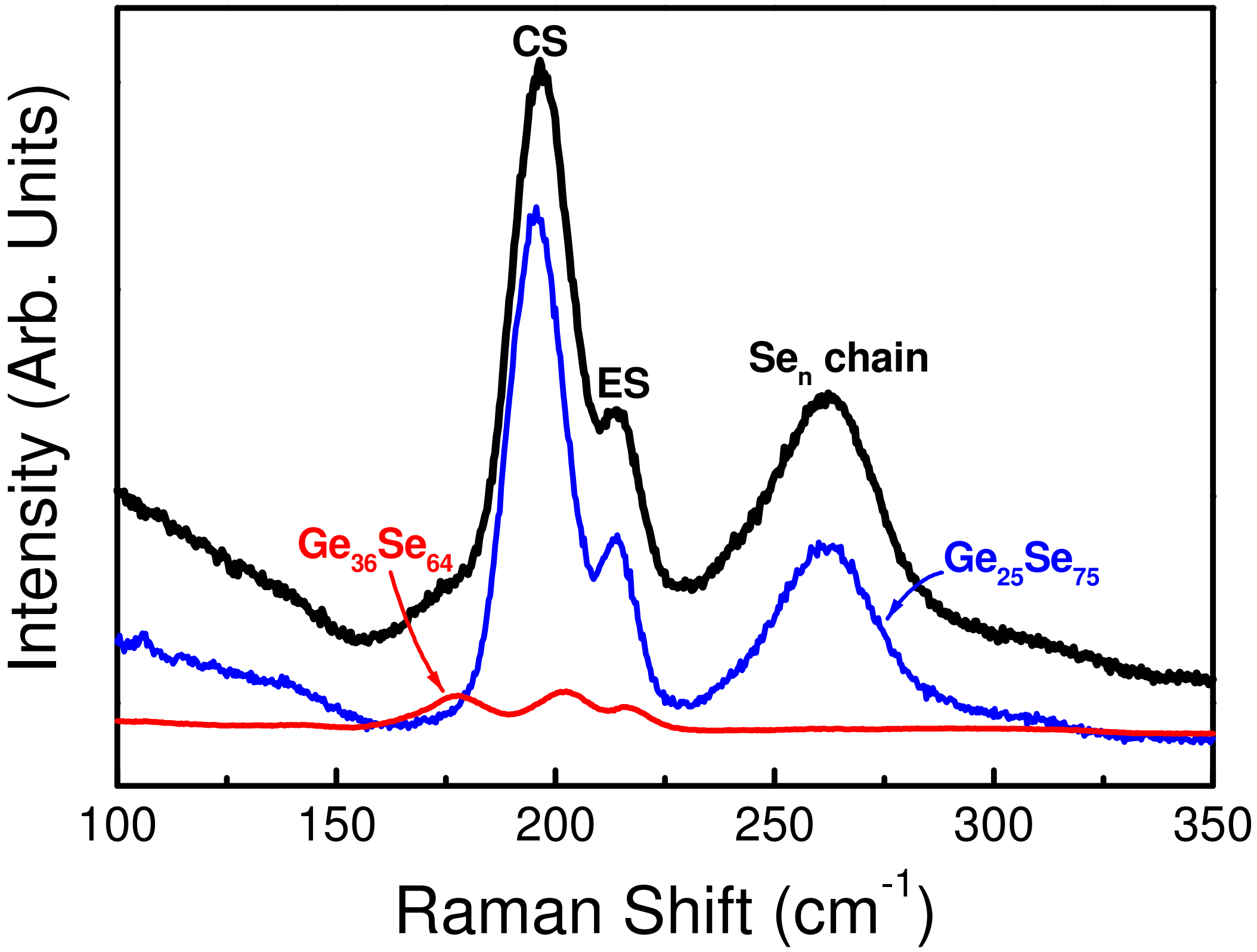

Figure 17

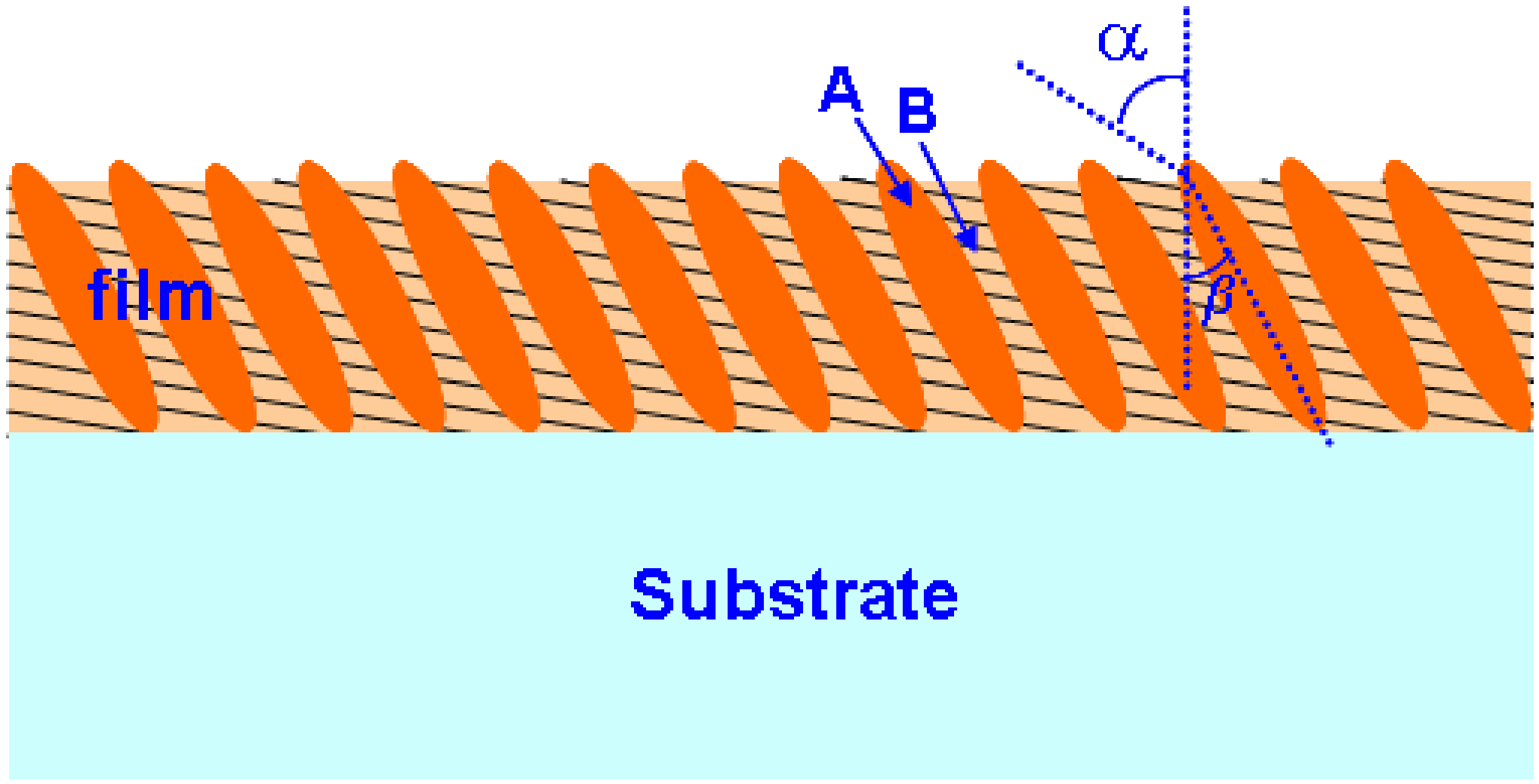

Figure 18